\documentstyle{article}
\expandafter\ifx\csname amssym.def\endcsname\relax \else\endinput\fi
\expandafter\edef\csname amssym.def\endcsname{%
       \catcode`\noexpand\@=\the\catcode`\@\space}
\catcode`\@=11
\def\undefine#1{\let#1\undefined}
\def\newsymbol#1#2#3#4#5{\let\next@\relax
 \ifnum#2=\@ne\let\next@\msafam@\else
\ifnum#2=\tw@\let\next@\msbfam@\fi\fi \mathchardef#1="#3\next@#4#5}
\def\mathhexbox@#1#2#3{\relax
 \ifmmode\mathpalette{}{\m@th\mathchar"#1#2#3}%
\else\leavevmode\hbox{$\m@th\mathchar"#1#2#3$}\fi}
\def\hexnumber@#1{\ifcase#1 0\or 1\or 2\or 3\or 4\or 5\or 6\or 7\or 8\or
 9\or A\or B\or C\or D\or E\or F\fi}

\ifcase\@ptsize
  \font\tenmsa=msam10 \font\sevenmsa=msam7 \font\fivemsa=msam5
\or
  \font\tenmsa=msam10 scaled \magstephalf \font\sevenmsa=msam7 scaled
\magstephalf \font\fivemsa=msam5 scaled \magstephalf
\or
  \font\tenmsa=msam10 scaled \magstep1 \font\sevenmsa=msam7 scaled
\fi

\newfam\msafam
\textfont\msafam=\tenmsa
\scriptfont\msafam=\sevenmsa
\scriptscriptfont\msafam=\fivemsa
\edef\msafam@{\hexnumber@\msafam}
\mathchardef\dabar@"0\msafam@39
\def\dashrightarrow{\mathrel{\dabar@\dabar@\mathchar"0\msafam@4B}}
\def\dashleftarrow{\mathrel{\mathchar"0\msafam@4C\dabar@\dabar@}}

\def\ulcorner{\delimiter"4\msafam@70\msafam@70 }
\def\urcorner{\delimiter"5\msafam@71\msafam@71 }
\def\llcorner{\delimiter"4\msafam@78\msafam@78 }
\def\lrcorner{\delimiter"5\msafam@79\msafam@79 }
\def\yen{{\mathhexbox@\msafam@55 }}
\def\checkmark{{\mathhexbox@\msafam@58 }}
\def\circledR{{\mathhexbox@\msafam@72 }}
\def\maltese{{\mathhexbox@\msafam@7A }}

\ifcase\@ptsize
  \font\tenmsb=msbm10 \font\sevenmsb=msbm7 \font\fivemsb=msbm5
\or
  \font\tenmsb=msbm10 scaled \magstephalf \font\sevenmsb=msbm7 scaled
\magstephalf \font\fivemsb=msbm5 scaled \magstephalf
\or
  \font\tenmsb=msbm10 scaled \magstep1 \font\sevenmsb=msbm7 scaled
\fi

\newfam\msbfam
\textfont\msbfam=\tenmsb
\scriptfont\msbfam=\sevenmsb
\scriptscriptfont\msbfam=\fivemsb
\edef\msbfam@{\hexnumber@\msbfam}
\def\Bbb#1{{\fam\msbfam\relax#1}}
\def\widehat#1{\setbox\z@\hbox{$\m@th#1$}%
 \ifdim\wd\z@>\tw@ em\mathaccent"0\msbfam@5B{#1}%
\else\mathaccent"0362{#1}\fi}
\def\widetilde#1{\setbox\z@\hbox{$\m@th#1$}%
 \ifdim\wd\z@>\tw@ em\mathaccent"0\msbfam@5D{#1}%
\else\mathaccent"0365{#1}\fi}

\ifcase\@ptsize
  \font\teneufm=eufm10 \font\seveneufm=eufm7 \font\fiveeufm=eufm5
\or
  \font\teneufm=eufm10 scaled \magstephalf \font\seveneufm=eufm7
scaled \magstephalf \font\fiveeufm=eufm5 scaled \magstephalf
\or
  \font\teneufm=eufm10 scaled \magstep1 \font\seveneufm=eufm7 scaled
\fi

\newfam\eufmfam
\textfont\eufmfam=\teneufm
\scriptfont\eufmfam=\seveneufm
\scriptscriptfont\eufmfam=\fiveeufm
\def\frak#1{{\fam\eufmfam\relax#1}}

\csname amssym.def\endcsname

\expandafter\ifx\csname pre amssym.tex at\endcsname\relax \else \endinput\fi
\expandafter\chardef\csname pre amssym.tex at\endcsname=\the\catcode`\@
\catcode`\@=11
\newsymbol\boxdot 1200
\newsymbol\boxplus 1201
\newsymbol\boxtimes 1202
\newsymbol\square 1003
\newsymbol\blacksquare 1004
\newsymbol\centerdot 1205
\newsymbol\lozenge 1006
\newsymbol\blacklozenge 1007
\newsymbol\circlearrowright 1308
\newsymbol\circlearrowleft 1309
\undefine\rightleftharpoons
\newsymbol\rightleftharpoons 130A
\newsymbol\leftrightharpoons 130B
\newsymbol\boxminus 120C
\newsymbol\Vdash 130D
\newsymbol\Vvdash 130E
\newsymbol\vDash 130F
\newsymbol\twoheadrightarrow 1310
\newsymbol\twoheadleftarrow 1311
\newsymbol\leftleftarrows 1312
\newsymbol\rightrightarrows 1313
\newsymbol\upuparrows 1314
\newsymbol\downdownarrows 1315
\newsymbol\upharpoonright 1316
 
\newsymbol\downharpoonright 1317
\newsymbol\upharpoonleft 1318
\newsymbol\downharpoonleft 1319
\newsymbol\rightarrowtail 131A
\newsymbol\leftarrowtail 131B
\newsymbol\leftrightarrows 131C
\newsymbol\rightleftarrows 131D
\newsymbol\Lsh 131E
\newsymbol\Rsh 131F
\newsymbol\rightsquigarrow 1320
\newsymbol\leftrightsquigarrow 1321
\newsymbol\looparrowleft 1322
\newsymbol\looparrowright 1323
\newsymbol\circeq 1324
\newsymbol\succsim 1325
\newsymbol\gtrsim 1326
\newsymbol\gtrapprox 1327
\newsymbol\multimap 1328
\newsymbol\therefore 1329
\newsymbol\because 132A
\newsymbol\doteqdot 132B
 
\newsymbol\triangleq 132C
\newsymbol\precsim 132D
\newsymbol\lesssim 132E
\newsymbol\lessapprox 132F
\newsymbol\eqslantless 1330
\newsymbol\eqslantgtr 1331
\newsymbol\curlyeqprec 1332
\newsymbol\curlyeqsucc 1333
\newsymbol\preccurlyeq 1334
\newsymbol\leqq 1335
\newsymbol\leqslant 1336
\newsymbol\lessgtr 1337
\newsymbol\backprime 1038
\newsymbol\risingdotseq 133A
\newsymbol\fallingdotseq 133B
\newsymbol\succcurlyeq 133C
\newsymbol\geqq 133D
\newsymbol\geqslant 133E
\newsymbol\gtrless 133F
\newsymbol\sqsubset 1340
\newsymbol\sqsupset 1341
\newsymbol\vartriangleright 1342
\newsymbol\vartriangleleft 1343
\newsymbol\trianglerighteq 1344
\newsymbol\trianglelefteq 1345
\newsymbol\bigstar 1046
\newsymbol\between 1347
\newsymbol\blacktriangledown 1048
\newsymbol\blacktriangleright 1349
\newsymbol\blacktriangleleft 134A
\newsymbol\vartriangle 134D
\newsymbol\blacktriangle 104E
\newsymbol\triangledown 104F
\newsymbol\eqcirc 1350
\newsymbol\lesseqgtr 1351
\newsymbol\gtreqless 1352
\newsymbol\lesseqqgtr 1353
\newsymbol\gtreqqless 1354
\newsymbol\Rrightarrow 1356
\newsymbol\Lleftarrow 1357
\newsymbol\veebar 1259
\newsymbol\barwedge 125A
\newsymbol\doublebarwedge 125B
\undefine\angle
\newsymbol\angle 105C
\newsymbol\measuredangle 105D
\newsymbol\sphericalangle 105E
\newsymbol\varpropto 135F
\newsymbol\smallsmile 1360
\newsymbol\smallfrown 1361
\newsymbol\Subset 1362
\newsymbol\Supset 1363
\newsymbol\Cup 1264
 
\newsymbol\Cap 1265
 
\newsymbol\curlywedge 1266
\newsymbol\curlyvee 1267
\newsymbol\leftthreetimes 1268
\newsymbol\rightthreetimes 1269
\newsymbol\subseteqq 136A
\newsymbol\supseteqq 136B
\newsymbol\bumpeq 136C
\newsymbol\Bumpeq 136D
\newsymbol\lll 136E
 
\newsymbol\ggg 136F
 
\newsymbol\circledS 1073
\newsymbol\pitchfork 1374
\newsymbol\dotplus 1275
\newsymbol\backsim 1376
\newsymbol\backsimeq 1377
\newsymbol\complement 107B
\newsymbol\intercal 127C
\newsymbol\circledcirc 127D
\newsymbol\circledast 127E
\newsymbol\circleddash 127F
\newsymbol\lvertneqq 2300
\newsymbol\gvertneqq 2301
\newsymbol\nleq 2302
\newsymbol\ngeq 2303
\newsymbol\nless 2304
\newsymbol\ngtr 2305
\newsymbol\nprec 2306
\newsymbol\nsucc 2307
\newsymbol\lneqq 2308
\newsymbol\gneqq 2309
\newsymbol\nleqslant 230A
\newsymbol\ngeqslant 230B
\newsymbol\lneq 230C
\newsymbol\gneq 230D
\newsymbol\npreceq 230E
\newsymbol\nsucceq 230F
\newsymbol\precnsim 2310
\newsymbol\succnsim 2311
\newsymbol\lnsim 2312
\newsymbol\gnsim 2313
\newsymbol\nleqq 2314
\newsymbol\ngeqq 2315
\newsymbol\precneqq 2316
\newsymbol\succneqq 2317
\newsymbol\precnapprox 2318
\newsymbol\succnapprox 2319
\newsymbol\lnapprox 231A
\newsymbol\gnapprox 231B
\newsymbol\nsim 231C
\newsymbol\ncong 231D
\newsymbol\diagup 231E
\newsymbol\diagdown 231F
\newsymbol\varsubsetneq 2320
\newsymbol\varsupsetneq 2321
\newsymbol\nsubseteqq 2322
\newsymbol\nsupseteqq 2323
\newsymbol\subsetneqq 2324
\newsymbol\supsetneqq 2325
\newsymbol\varsubsetneqq 2326
\newsymbol\varsupsetneqq 2327
\newsymbol\subsetneq 2328
\newsymbol\supsetneq 2329
\newsymbol\nsubseteq 232A
\newsymbol\nsupseteq 232B
\newsymbol\nparallel 232C
\newsymbol\nmid 232D
\newsymbol\nshortmid 232E
\newsymbol\nshortparallel 232F
\newsymbol\nvdash 2330
\newsymbol\nVdash 2331
\newsymbol\nvDash 2332
\newsymbol\nVDash 2333
\newsymbol\ntrianglerighteq 2334
\newsymbol\ntrianglelefteq 2335
\newsymbol\ntriangleleft 2336
\newsymbol\ntriangleright 2337
\newsymbol\nleftarrow 2338
\newsymbol\nrightarrow 2339
\newsymbol\nLeftarrow 233A
\newsymbol\nRightarrow 233B
\newsymbol\nLeftrightarrow 233C
\newsymbol\nleftrightarrow 233D
\newsymbol\divideontimes 223E
\newsymbol\varnothing 203F
\newsymbol\nexists 2040
\newsymbol\Finv 2060
\newsymbol\Game 2061
\newsymbol\mho 2066
\newsymbol\eth 2067
\newsymbol\eqsim 2368
\newsymbol\beth 2069
\newsymbol\gimel 206A
\newsymbol\daleth 206B
\newsymbol\lessdot 236C
\newsymbol\gtrdot 236D
\newsymbol\ltimes 226E
\newsymbol\rtimes 226F
\newsymbol\shortmid 2370
\newsymbol\shortparallel 2371
\newsymbol\smallsetminus 2272
\newsymbol\thicksim 2373
\newsymbol\thickapprox 2374
\newsymbol\approxeq 2375
\newsymbol\succapprox 2376
\newsymbol\precapprox 2377
\newsymbol\curvearrowleft 2378
\newsymbol\curvearrowright 2379
\newsymbol\digamma 207A
\newsymbol\varkappa 207B
\newsymbol\Bbbk 207C
\newsymbol\hslash 207D
\undefine\hbar
\newsymbol\hbar 207E
\newsymbol\backepsilon 237F
\catcode`\@=\csname pre amssym.tex at\endcsname

\def\Box{\hbox{\vrule height1ex\kern-0.4pt
\vbox to 1ex{\hrule width1ex\vfil\hrule width1ex}\kern-0.4pt\vrule height1ex}}
\newcommand{\sqr}[2]{{{\vcenter{\vbox{\hrule height.#2pt
\hbox{\vrule width.#2pt height#1pt \kern#1pt
\vrule width.#2pt}
\hrule height.#2pt}}}}}

\newtheorem{proposition}{Proposition}

\newcommand{\ovl}{\overline}

\newtheorem{definition}{Definition}

\newcommand{\eo}{\setcounter{equation}{0}}

\newcommand{\be}{\begin{equation}}

\newcommand{\ee}{\end{equation}}
\newcommand{\al}{\alpha}
\newcommand{\bt}{\beta}
\newcommand{\gm}{\gamma}
\newcommand{\dl}{\delta}

\newcommand{\varep}{\varepsilon}

\newcommand{\th}{\theta}

\newcommand{\lm}{\lambda}

\newcommand{\rh}{\rho}
\newcommand{\sg}{\sigma}
\newcommand{\ta}{\tau}

\newcommand{\phv}{\varphi}

\newcommand{\Sg}{\Sigma}
\newcommand{\ps}{\psi}
\newcommand{\Ps}{\Psi}
\newcommand{\om}{\omega}
\newcommand{\Om}{\Omega}

\newcommand{\raw}{\rightarrow}
\newcommand{\A}{\frak A}
\newcommand{\B}{\frak B}

\newcommand{\C}{{\Bbb C}}
\newcommand{\ca}{$C^*$-algebra}
\newcommand{\jba}{$JB$-algebra}
\newcommand{\rep}{representation}
\newcommand{\bib}{\bibitem}
\newcommand{\cin}{C^{\infty}}
\newcommand{\M}{{\frak M}}
\renewcommand{\H}{\mbox{$\cal H$}}
\newcommand{\KH}{{\frak K}({\cal H})}
\newcommand{\BH}{{\frak B}({\cal H})}
\newcommand{\PH}{{\Bbb P}{\cal H}}
\newcommand{\n}{\parallel}

 \renewcommand{\ll}{\label}
\newcommand{\Ac}{{\frak A}_{\Bbb C}} 
 \newcommand{\R}{{\Bbb R}}
\newcommand{\ci}{\cite}
\newcommand{\bea}{\begin{eqnarray}}
\newcommand{\eea}{\end{eqnarray}}

\newcommand{\half}{\mbox{\footnotesize $\frac{1}{2}$}}
\newcommand{\quar}{\mbox{\footnotesize $\frac{1}{4}$}}
\newcommand{\D}{{\Bbb D}}

\topmargin = - 0.5 cm
\newcommand{\enp}{\hfill $\blacksquare$}
\newcommand{\la}{\langle}
\newcommand{\ra}{\rangle}
\renewcommand{\P}{{\cal P}}
\renewcommand{\A}{{\frak A}}
\newcommand{\eb}{\partial_e K}
\newcommand{\wed}{\wedge}
\newcommand{\AOP}{{\frak A}_0({\cal P})}
\newcommand{\AP}{{\frak A}({\cal P})}
\newcommand{\tps}{transition probability space}
\newcommand{\tp}{transition probability}
\newcommand{\tpies}{transition probabilities} 
\newcommand{\Hs}{Hilbert space}

\newcommand{\lp}{{\cal L}({\cal P})}
\renewcommand{\L}{{\cal L}}
\newcommand{\LP}{{\cal L}({\cal P})}
\newcommand{\PA}{{\cal P}({\frak A})}
\newcommand{\SA}{{\cal S}({\frak A})}
\begin{document}
\setlength{\baselineskip}{1\baselineskip}
\thispagestyle{empty}
\title{Poisson spaces with a transition probability}
\author{ N.P.~Landsman\thanks{E.P.S.R.C. Advanced Research Fellow}
 \\ \mbox{}\hfill \\ Department of Applied Mathematics and Theoretical
Physics\\ University of Cambridge, Silver Street, Cambridge CB3 9EW,
U.K. } \maketitle
\begin{abstract}
The common structure of the space of pure states $\P$ of a classical
or a quantum mechanical system is that of a Poisson space with a
transition probability. This is a topological space equipped with a
Poisson structure, as well as with a function $p:\P\times\P\raw
[0,1]$, with certain properties. The Poisson structure is connected
with the transition probabilities through unitarity (in a specific
formulation intrinsic to the given context).

In classical mechanics, where $p(\rh,\sg)=\dl_{\rh\sg}$, unitarity
poses no restriction on the Poisson structure. Quantum mechanics is
characterized by a specific (complex Hilbert space) form of $p$, and
by the property that the irreducible components of $\P$ as a
transition probability space coincide with the symplectic leaves of
$\P$ as a Poisson space.  In conjunction, these stipulations determine
the Poisson structure of quantum mechanics up to a multiplicative
constant (identified with Planck's constant).

Motivated by E.M. Alfsen, H. Hanche-Olsen and F.W. Shultz ({\em Acta
Math.} {\bf 144} (1980) 267-305) and F.W. Shultz ({\em Commun.\ Math.\
Phys.} {\bf 82} (1982) 497-509), we give axioms guaranteeing that $\P$
is the space of pure states of a unital $C^*$-algebra. We give an
explicit construction of this algebra from $\P$.
\end{abstract}
 \newpage
\section{Introduction}
Section 1.1 motivates the axiomatic study of state spaces (rather than
operator algebras) in the foundations of quantum mechanics. In 1.2 we
review the work of Alfsen et al.\ on the structure of state spaces of
$C^*$-algebras.  In 1.3 we discuss the concept of a \tps, and in 1.4
it is shown how the pure state space of a \ca\ is an example of such a
space. Section 1.5 recalls the concept of a Poisson manifold, and
introduces (uniform) Poisson spaces generalizing this concept. Poisson
structures may be intertwined with
\tpies, leading to the notion of unitarity, 
and to the central idea of this paper, a Poisson space
with a \tp.

In Chapter 2 we introduce our axioms on pure state spaces, and
formulate the theorem relating these axioms to pure state spaces of
\ca s. Chapter 3 outlines the proof of this theorem, which essentially
consists of the reconstruction of a \ca\ from its pure state space,
endowed with the structure of a uniform Poisson space with a \tp.
This reconstruction is of interest in its own right.  Some longer
proofs and other technical comments appear in Chapter 4.
 
 {\em In this paper functions and functionals are real-valued, unless
explicitly indicated otherwise. Hence $C(X)$ stands for $C(X,\R)$,
etc. Similarly, vector spaces (including the various algebras
appearing in this paper) are generally over $\R$, unless there is an
explicit label $\C$ denoting complexification.  An exception to this
rule is that we use the standard symbols $\H$ for a complex Hilbert
space, and $\BH$ ($\KH$) for the set of all bounded (compact)
operators on $\H$.  The self-adjoint part of a \ca\ $\Ac$ is denoted
by $\A$; we denote the state space of $\Ac$ by $\SA$ or ${\cal
S}(\Ac)$, and its pure state space by $\PA$ or ${\cal P}(\Ac)$.  Here
the `pure state space' is the space of all pure states, rather than
its $w^*$-closure.} \subsection{Algebraic aspects of mechanics} At
face value, quantum mechanics (Hilbert space, linear operators) looks
completely different from classical mechanics (symplectic manifolds,
smooth functions).  The structure of their respective algebras of
observables, however, is strikingly similar.  In quantum mechanics,
one may assume \ci{RR,Emc} that the observables $\A$ form the
self-adjoint part of some $C^*$-algebra $\A_{\C}$. The associative
product does not map $\A$ into itself, but the anti-commutator $A\circ
B=\half(AB+BA)$ and the (scaled) commutator
$[A,B]_{\hbar}=i(AB-BA)/\hbar$ do; in conjunction, they give $\A$ the
structure of a so-called {\bf Jordan-Lie algebra} \ci{GP,Emc}.  This
is a vector space $V$ equipped with two bilinear maps $\circ$ and
$[\;,\;]:V\times V\raw V$, such that $\circ$ is symmetric, $[\;,\;]$
is a Lie bracket (i.e., it is anti-symmetric and satisfies the Jacobi
identity), and the Leibniz property
\be
[A,B\circ C]=[A,B]\circ C+ B\circ [A,C] \ll{leib}
\ee
 holds; in other words, the commutator is a derivation of the Jordan
product. Moreover, one requires the associator identity \be (A\circ
B)\circ C-A\circ (B\circ C)= k[[A,C],B] \ll{ass}
\ee
for some $k\in\R$. This implies the Jordan identity $A^2\circ (A\circ
B)=A\circ (A^2\circ B)$ (where $A^2=A\circ A$), which makes
$(V,\circ)$ a Jordan algebra \ci{Emc,HOS}); accordingly, the symmetric
product $\circ$ is referred to as the {\bf Jordan product}.  Note that
for $V=\A$ and $[A,B]=[A,B]_{\hbar}$ one has $k=\hbar^2/4$.

Conversely, a Jordan-Lie algebra $\A$ for which $k>0$ (cf.\ \ci{Emc}
for comments on the case $k<0$), and which in addition is a so-called
$JB$-algebra, is the self-adjoint part of a $C^*$-algebra $\A_{\C}$.

Here a {\bf $JB$-algebra} \ci{ASS,HOS} is a Jordan algebra which is a
Banach space, and satisfies $\n A\circ B\n\, \leq\, \n A\n \, \n B\n$,
$\n A^2\n =\n A\n^2$, and $\n A^2\n\, \leq\,
\n A^2+ B^2\n$ for all $A,B\in \A$; 
the first axiom can actually be derived from the other two;
alternatively, the last two axioms may be replaced by $\n A\n^2\,
\leq\,
\n A^2+ B^2\n$. 

The associative $C^*$-product is given by $A\cdot B=A\circ B
-i\sqrt{k}[A,B]$ (the $\cdot$ is usually omitted); the associativity
follows from the Leibniz property, (\ref{ass}), and the Jacobi
identity.  For the construction of the norm and the verification of
the axioms for a $C^*$-algebra, see \ci{Wri,Rod} and section 3.8
below.

In classical mechanics, one takes the Jordan-Lie algebra to consist of
all smooth functions on the phase space, equipped with the operations
of pointwise multiplication $f\circ g=fg$ and Poisson bracket
$[f,g]=\{f,g\}$ (the latter coming from a symplectic structure, or
from a more general abstract Poisson structure
\ci{Wei,MR}). The identity (\ref{ass}) is then satisfied with
$k=0$. A Jordan-Lie algebra for which $k=0$ in (\ref{ass}) is called a
{\bf Poisson algebra}.

Thus from an algebraic point of view the only difference between
classical and quantum mechanics is that in the former the Jordan
product $\circ$ is associative, whereas in the latter the more general
identity (\ref{ass}) is satisfied for some $k>0$.

From an axiomatic point of view, it is rather difficult to justify
(\ref{ass}), and it is hard to swallow that the non-associativity of
$\circ$ should be the defining property of quantum mechanics.
Historically, the commutator hardly played a role in algebraic quantum
axiomatics, all attention being focused on the Jordan structure
\ci{vN1,Seg,ASS,HOS,Emc}.  Whereas the Jordan identity may be
justified by the need to have a spectral theory, the step from the
Jordan- to the full $C^*$-structure has had to be justified
algebraically by an appeal to the need to combine different physical
systems using a well-behaved tensor product \ci{Ara,HO}.  This gives
the commutator a different status from its classical counterpart
(viz.\ the Poisson bracket), which describes the way observables lead
to flows (i.e., dynamics).
\subsection{State spaces and the work of Alfsen, Shultz, and Hanche-Olsen}
A transparent way of analyzing and justifying algebras of observables
is the study of their state spaces. A state on a \jba\ $\A$ is defined
as a linear functional $\om$ on $\A$ satisfying $\om(A^2)\geq 0$ for
all $A\in\A$ and $\n \om
\n=1$; in case that $\A$ has an identity $\Bbb I$ this implies that
$\om({\Bbb I})=1$. The idea is that the algebraic structure of $\A$ is
encoded in certain (geometric) properties of its state space ${\cal
S}(\A)$, so that $\A$ may be reconstructed from ${\cal S}(\A)$,
equipped with these properties. The most basic property of ${\cal
S}(\A)$ is that it is a convex set, which is compact in the
$w^*$-topology if $\A$ is a $JB$-algebra with unit.  The description
of quantum mechanics in terms of general compact convex state spaces
is closely tied to the so-called operational approach, and is
invariably interpreted in terms of laboratory procedures such as
filtering measurements \ci{Sch,Mie1,Mie2,Mie3,Lud,BC,Kum}.

For $C^*$-algebras (which are special instances of complexified
$JB$-algebras) this type of study culminated in
\ci{AHS}, where axioms were given which guarantee that
 a given compact convex set $K$ (assumed to be embedded in a locally
convex Hausdorff vector space) is the state space of a $C^*$-algebra
with unit (also cf.\ \ci{Alf2,AE,AS3}). In order to motivate our own
approach, we need to explain these axioms to some extent.

Firstly, a {\bf face} $F$ is defined as a convex subset of $K$ with
the property that $\rh$ and $\sg$ are in $F$ if $\lm\sg+ (1-\lm)\rh\in
F$ for some $\lm\in (0,1)$.  A face $F$ is called {\bf norm-exposed}
\ci{AS2} if it equals $F=\{\rh\in K|\la f,\rh\ra=0\}$ for some $f\in
A^+_b(K)$.  Here $A_b(K)$ is the space of all bounded affine functions
on $K$, and $A^+_b(K)$ its subspace of positive functions. $A(K)$ will
stand for the space of continuous affine functions on $K$
\ci{AS1,AE}.

A face $F$ is said to be {\bf projective} \ci{AS1} if there exists
another face $F^{\sharp}$ such that $F$ and $F^{\sharp}$ are
norm-exposed and affinely independent \ci{Alf1}, and there exists a
map (a so-called affine retraction) $\ps:K\raw K$ with image the
convex sum of $F$ and $F^{\sharp}$, leaving its image pointwise
invariant, and having the technical property of transversality (cf.\
\ci[3.8]{AS1} or \ci{Alf2}) (alternative definitions are possible
\ci{AS1}).
 The first axiom of
\ci{AHS} is

{\bf Axiom AHS1}: {\em Every norm-exposed face of $K$ is projective.}
\\ A face consisting of one point is called a pure state, and the
collection of pure states forms the so-called {\bf extreme boundary}
$\partial_e K$ of $K$. The smallest face containing a subset $S\subset
K$ is denoted by $F(S)$, and we write $F(\rh,\sg)$ for
$F(\{\rh,\sg\})$. Two pure states $\rh,\sg$ are called inequivalent if
$F(\rh,\sg)$ is the line segment $\{\lm\sg+ (1-\lm)\rh|\lm\in
[0,1]\}$. Otherwise, they are called {\bf equivalent}. The second
axiom is

 {\bf Axiom AHS2}: {\em If pure states $\rh$ and $\sg\neq \rh$ are
equivalent, then $F(\rh,\sg)$ is norm-exposed and affinely isomorphic
to the state space of the $C^*$-algebra $\M_2(\C)$ of $2\times 2$
matrices over $\C$. Moreover, each pure state is norm-exposed.}

The state space ${\cal S}(\M_2(\C))$ is affinely isomorphic to the
unit ball $B^3$ in $\R^3$.  Concretely, we identify a state on
$\M_2(\C)$ with a density matrix on $\C^2$, which may be parametrized
as
\be
\half 
\left(
\begin{array}{cc}
1 + x & y+iz \\ y-iz & 1 -x
\end{array}
\right) , \ll{parrho}
\ee
where $x,y,z\in\R$. The positivity of this matrix then corresponds to
the constraint $x^2 + y^2 + z^2\leq 1$ (see \ci{AHS}).

From the point of view of quantum logic (cf.\ e.g.\ \ci{Var,BC,Kal1}),
Axiom AHS1 allows one to define an orthomodular lattice, whose
elements are the projective faces of $K$ \ci[\S 4]{AS1}. Axiom AHS2
not only allows one to prove that this lattice has the covering
property \ci[6.15]{AS3}, but also eventually implies that the
co-ordinatizing field of the lattice is $\C$ (cf.\ section 4.1).  In
the finite-dimensional case Axiom AHS1 and AHS2 are sufficient to
construct a $C^*$-algebra $\A_{\C}$ whose state space is $K$; as a
Banach space $\A=A(K)$ with the sup-norm.  To cover the general case,
more axioms are needed.

{\bf Axiom AHS3}: {\em The $\sg$-convex hull of $\partial_e K$ is a
split face of $K$.}\\ Here the {\bf $\sg$-convex hull} in question
consist of all sums $\sum_i \lm_i\rh_i$, where $\rh_i\in
\partial_e K$,
$\lm_i\in[0,1]$, $\sum_i\lm_i=1$, and the sum converging in the norm
topology (regarding $K$ as a subset of the dual of the Banach space
$A(K)$). A face $F$ of $K$ is {\bf split} if there exists another face
$F'$ such that $K=F\oplus_c F'$ (direct convex sum).  Let $C\subset
\partial_eK$ consist of all pure states in a given equivalence class,
and let $\overline{F}(C)$ be the $\sg$-convex hull of $C$ (this
coincides with the smallest split face containing any member of $C$).
Then $A_b(\overline{F}(C))_{\C}$ can be made into a von Neumann
algebra (with predual $\overline{F}(C)_{\C}$) on the basis of axioms
1-3
\ci[\S 6]{AS3}, \ci[\S 6]{AHS}.
 Axiom AHS3 is used to show that this is an atomic (type I) factor,
i.e., ${\frak B}(\H_C)$ for some Hilbert space $\H_C$.

The remaining axioms serve to combine all the $A(\overline{F}(C))$
into $A(K)$ in such a way that one obtains the self-adjoint part of a
$C^*$-algebra. The Jordan product $A\circ B$ (or, equivalently, $A^2$)
is constructed using the non-commutative spectral theory defined by
$K$ \ci{AS1,AS2}. This product then coincides with the anti-commutator
in $A_b(\overline{F}(C))\simeq {\frak B}(\H_C)$.  In principle this
could map $A\in A(K)$ into $A^2\in A_b(K)$ (that is, not necessarily
in $A(K)$). Hence

 {\bf Axiom AHS4}: {\em if $A\in A(K)$ then $A^2\in A(K)$.}\\ This is
not the formulation of the axiom given in \ci{AS3,AHS}, but by
\ci[9.6]{AS1},
\ci[7.2]{AS3} it is immediately equivalent to the version
 in the literature. Finally,
the commutator, already defined on each $A(\overline{F}(C))$, needs to
be well-defined on all of $A(K)$. This is guaranteed by

 {\bf Axiom AHS5}: {\em $K$ is orientable.}\\ Roughly speaking, this
means that one cannot transport a given face $F(\rh,\sg)\simeq B^3$
(cf.\ Axiom AHS2) in a continuous way around a closed loop so that is
changes its orientation (cf.\ \ci[\S 7]{AHS} for more detail; also
cf.\ section 4.3 below). It is remarkable that $A(K)$ is automatically
closed under the commutator, given the axioms.  It is proved in
\ci{AHS} that a compact convex set is the state space of a unital
$C^*$-algebra iff Axioms AHS1-AHS5 are satisfied.
 
Even if one is happy describing quantum mechanics with superselection
rules in terms of $C^*$-algebras, from a physical perspective one
should not necessarily regard the above axioms as unique, or as the
best ones possible.  The notion of a projective face (or,
equivalently, a $P$-projection \ci{AS1}) is a complicated one (but
cf.\ \ci{Ara} for a certain simplification in the finite-dimensional
case, and \ci{Kum} for an analogous interpretation in terms of filters
in the general case). One would like to replace the concept of
orientability by some statement of physical appeal. Most importantly,
the comparison of classical and quantum mechanics seems facilitated if
one could start from the space of pure states $\partial_e K$ as the
basic object. Moreover, from an ontological rather than an
epistemological point of view one would prefer a formulation in terms
of pure states as well, and the same comment applies if one is
interested in an individual (as opposed to a statistical)
interpretation of quantum mechanics.  \subsection{Transition
probability spaces} Clearly, the extreme boundary $\eb$ of a given
compact convex set $K$ as a topological space does not contain enough
information to reconstruct $K$. However, one can equip $\eb$ with the
additional structure of a so-called transition probability, as first
indicated by Mielnik \ci{Mie2} (also cf.\ \ci{Shu}).  Namely, given
$\rh,\sg\in\eb$ one can define $p$ by
\be 
p(\rh,\sg)={\rm inf} \{f(\rh)| f\in A_b(K), 0\leq f\leq
1,f(\sg)=1\}.\ll{mtp}
\ee
For later use, we notice that it follows that
\be 
p(\sg,\rh)=1-{\rm sup} \{f(\sg)| f\in A_b(K), 0\leq f\leq
1,f(\rh)=0\}.\ll{mtpalter}
\ee

For the moment we denote $\eb$ by $P$. By construction,
\be
p:\P\times\P\raw[0,1] \ll{tp1}
\ee
satisfies $\rh=\sg\Rightarrow p(\rh,\sg)=1$.  Moreover, we infer from
(\ref{mtpalter}) that
\be
p(\rh,\sg)=0 \, \Longleftrightarrow \, p(\sg,\rh)=0. \ll{tp2half}
\ee
If $K$ has the property that every pure state is norm-exposed, then,
as is easily verified, $p(\rh,\sg)=1\Rightarrow\rh=\sg$, so that
\be
p(\rh,\sg)=1 \, \Longleftrightarrow \,\rh=\sg. \ll{tp2}
\ee

Any function $p$ on a set $\P$ with the properties (\ref{tp1}),
(\ref{tp2half}), and (\ref{tp2}) is called a {\bf transition
probability}, and $\P$ is accordingly called a {\bf transition
probability space}. (In its abstract form these concepts are due to
von Neumann \ci{vN2}, who in addition required $p$ to satisfy
(\ref{tp3}) below; also cf.\ \ci{Mie1,Zab,Bel,BC,Pul1}). A transition
probability is called {\bf symmetric} if \be p(\rh,\sg)=p(\sg,\rh)\:\:
\forall \rh,\sg\in\P. \ll{tp3}
\ee

A subset $S\subset\P$ is called orthogonal if $p(\rh,\sg)=0$ for all
pairs $\rh\neq\sg$ in $S$. A basis $B$ of $\P$ is an orthogonal subset
for which $\sum_{\rh\in B}p(\rh,\sg)=1$ for all $\sg\in\P$ (here the
sum is defined as the supremum of all finite partial sums).  A basic
theorem is that all bases of a given symmetric \tps\ have the same
cardinality \ci{Mie1}; this cardinality is the {\bf dimension} of
$\P$.

 One imposes the requirement
\be 
\mbox{\em Every maximal  orthogonal subset of $\P$  is  a  basis}.
\ll{tp4} 
\ee

A \tps\ is called {\bf irreducible} if it is not the union of two
(nonempty) orthogonal subsets.  A {\bf component} $C$ is a subset of
$\P$ with the property that $p(\rh,\sg)=0$ for all $\rh\in C$ all
$\sg\in\P\backslash C$. Thus a \tps\ is the disjoint union of its
irreducible components \ci{Bel}.  An irreducible component of $\P$ is
called a {\bf sector}. This agrees with the terminology in algebraic
quantum mechanics, where $\P$ is the pure state space of a
$C^*$-algebra (of observables) \ci{RR}. If one defines a topology on
$\P$ through the metric $d(\sg,\rh)={\rm l.u.b.}
\{|p(\rh,\ta)-p(\sg,\ta)|, \ta\in\P\}$ \ci{Bel}, then the topological
components coincide with the components just defined. However, a
different topology may be defined on $\P$, and therefore we shall use
the term `sector' as referring to `component' in the first
(probabilistic) sense. Two points lying in the same sector of $\P$ are
called {\bf equivalent} (and {\bf inequivalent} in the opposite case).

Any subset $Q\subset \P$ has an orthoplement
$Q^{\perp}=\{\sg\in\P|p(\rh,\sg)=0\: \forall \rh\in Q\}$.  One always
has $Q\subseteq Q^{\perp\perp}$; a subset $Q$ is called {\bf
orthoclosed} if $Q=Q^{\perp\perp}$. Any set of the type $ Q^{\perp}$
(hence in particular $Q^{\perp\perp}$) is orthoclosed. In particular,
one may choose a orthogonal subset $S$, in which case \ci{Mie1,Zab}
$S^{\perp\perp}=\{\rh\in\P|\sum_{\sg\in S} p(\rh,\sg)=1\}$. (Clearly,
if $S=B$ is a basis then $B^{\perp\perp}=\P$.) Not every orthoclosed
subset is necessarily of this form, however: there exist examples of
orthoclosed subsets which do not have any basis \ci{Zab,BC}. To
exclude pathological cases, one therefore adds the axiom \ci{Zab,BC}
\be
\mbox{\em If $Q\subseteq\P$ is orthoclosed then
 every maximal orthogonal subset of $Q$ is a basis
of $Q$.}
\ll{tp5} 
\ee
\begin{definition} 
 A {\bf well-behaved \tps} is a pair $(\P,p)$ satisfying
(\ref{tp1})-(\ref{tp5}).\ll{defwbtps}
\end{definition} 
Of course, (\ref{tp2half}) and (\ref{tp4}) follow from (\ref{tp3}) and
(\ref{tp5}), respectively.

The simplest example of a well-behaved \tps\ is given by putting the
`classical' \tpies
\be
p(\rh,\sg)=\dl_{\rh\sg} \ll{tpcm}
\ee
on any set $\P$.

One can associate a certain function space $\AP$ with any \tps\ $\P$.
Firstly, for each $\rh\in\P$ define $p_{\rh}\in\ell^{\infty}(\P)$ by
\be
p_{\rh}(\sg)=p(\rh,\sg). \ll{defprho}
\ee
Secondly, the normed vector space $\A_{00}(\P)$, regarded as a
subspace of $\ell^{\infty}(\P)$ (with sup-norm), consists of all
finite linear combinations of the type $\sum_{i=1}^N c_i p_{\rh_i}$,
where $c_i\in\R$ and $\rh_i\in\P$. The closure of $\A_{00}(\P)$ is
called $\A_0(\P)$.  Thirdly, the double dual of $\AOP$ will play a
central role in what follows, so that we use a special symbol: \be
\AP=\A_0(\P)^{**}. \ll{defap}
\ee
Since $\AOP\subseteq \ell_0(\P)$, one has $\AP\subseteq
\ell_0(\P)^{**}=\ell^{\infty}(\P)$.  The space $\AP$ is the function
space intrinsically related to a \tps\ $\P$. In the case (\ref{tpcm})
one immediately finds $\AP=\ell^{\infty}(\P)$.

(Following a seminar the author gave in G\"{o}ttingen, 1995, A.
Uhlmann informed him that in his lectures on quantum mechanics
$\A_{00}(\P)$ had long been employed as the space of observables.)
\subsection{Transition probabilities on pure state spaces}
Using the results in \ci{AS3} (in particular, the so-called `pure
state properties') as well as Thm.\ 2.17 in \ci{AS1}, it is not
difficult to show that the pure state space of a unital $JB$-algebra
(where every pure state is indeed norm-exposed) is a symmetric
transition probability space.

If one further specializes to the pure state space $\PA$ of a unital
$C^*$-algebra $\Ac$, from (\ref{mtp}) one may derive the explicit
expression \be p(\rh,\sg)=1-{\mbox{\footnotesize $\frac{1}{4}$}}\n
\rh-\sg\n^2,
\ll{tpinca}
\ee
which coincides with
\be 
p(\rh,\sg)=|(\Om_{\rh},\Om_{\sg})|^2 \ll{tpsforca}
\ee
 if $\rh$ and $\sg$ are equivalent (where $\Om_{\rh}$ is a unit vector
implementing $\rh$ in the corresponding GNS representation, etc.), and
equals 0 if they are not; cf.\ \ci{GK,RR,Shu}. This will be proved in
section 4.2.

The notion of equivalence between pure states used here may refer
either to the one defined between eqs.\ (\ref{tp4}) and (\ref{tp5}) in
the context of \tps s, or to the unitary equivalence of the
GNS-representations defined by the states in question in the context
of $C^*$-algebras; these notions coincide. In fact, $\PA$ has the
following decomposition into sectors (see \ci{RR}, which on this point
relies on \ci{GK}):
\be
\P(\A)=\cup_{\al}\PH_{\al},\ll{pssdec}
\ee
 where $\H_{\al}$ is isomorphic to the irreducible GNS-\rep\ space of
an arbitrary state in the projective Hilbert space $\PH_{\al}$.  All
states in a given subspace $\PH_{\al}$ are equivalent, and any two
states lying in different such subspaces are inequivalent.

 We regard the self-adjoint part $\A$ of $\Ac$ as a subspace of
$C(\PA)$ (where $\PA$ is equipped with the $w^*$-topology inherited
from $\SA$) through the Gel'fand transform $\hat{A}(\rh)=\rh(A)$, for
arbitrary $A\in\A$ and $\rh\in\PA$. Similary, an operator $A\in\BH$ is
identified with a function $\hat{A}\in C(\PH)$ through the canonical
inclusion $\PH\subset {\cal S}(\BH)$ (where $\PH$ carries the
$w^*$-topology relative to this inclusion).  Under these
identifications, for each $\rh\in\PA$ the irreducible representation
$\pi_{\rh}(\A)$ is unitarily equivalent to the restriction of $\A$ to
the sector containing $\rh$; every irreducible representation of $\A$
is therefore given (up to unitary equivalence) by the restriction of
$\A$ to one of its sectors.

 In any case, one recovers the usual transition probabilities of
quantum mechanics.  If $\Ac=\KH$ (or $\M_N(\C)=\B(\C^N)$), the pure
state space $\P(\KH)$ is the projective Hilbert space $\PH$ (or ${\Bbb
P}\C^N$). One may then equally well interpret $\Om_{\rh}$ (etc.) in
(\ref{tpsforca}) as a lift of $\rh\in\PH$ to the unit sphere $\Bbb
S\H$ in $\H$.

In particular, it follows that the pure state space of a unital
$C^*$-algebra is a well-behaved transition probability space.  The
space $\A(\PA)$ can be explicitly identified.  Let $\pi_{\rm ra}$ be
the reduced atomic representation of $\Ac$ \ci{KR1}; recall that
$\pi_{\rm ra}$ is the direct sum over irreducible \rep s $\pi_{\rm
ra}=\oplus_{\rh} \pi_{\rh}$ (on the \Hs\ $\H_{\rm ra}=
\oplus_{\rh} \H_{\rh}$), where one
includes one representative of each equivalence class in $\P(\A)$.
For the weak closure one obtains $\pi_{\rm ra}(\Ac)^-=\oplus_{\rh}
\B(\H_{\rh})$. The Gel'fand transform maps $\pi_{\rm ra}(\A)^-$ into a
subspace of $\ell^{\infty}(\PA)$. It will be shown in section
\ref{aopdd} that this subspace is precisely $\A(\PA)$; we write this as
\be
\A(\PA)=\hat{\pi}_{\rm ra}(\A)^-.  \ll{oldapaispira}
\ee
The isomorphism between $\pi_{\rm ra}(\A)^-$ and $\A(\PA)$ thus
obtained is isometric and preserves positivity (since the Gel'fand
transform does).

For any well-behaved transition probability space $\P$ one can define
a lattice $\lp$, whose elements are the orthoclosed subsets of $\P$
(including the empty set $\emptyset$, and $\P$ itself).  The lattice
operations are: $Q\leq R$ means $Q\subseteq R$, $Q\wedge R=Q\cap R$,
and $Q\vee R=(Q\cup R)^{\perp\perp}$.  The zero element $0$ is
$\emptyset$. Note that the dimension of $\lp$ as a lattice equals the
dimension \ci{Kal1} of $\P$ as a \tps. It is orthocomplemented by
$\perp$, and is easily shown to be a complete atomic orthomodular
lattice \ci{Zab,Bel,BC} (cf.\ \ci{Kal1} for the general theory of
orthomodular lattices). In our approach, this lattice plays a somewhat
similar role to the lattice ${\cal F}(K)$ of projective faces of $K$
(or, equivalently, of $P$-projections \ci{AS1}; note that for
$C^*$-algebras ${\cal L}(\partial_e K)$ is not necessarily isomorphic
to ${\cal F}(K)$).  However, it seems to us that both the definition
and the physical significance of $\lp$ are more direct.

If $\P$ is a classical \tps\ (see \ref{tpcm}) then $\lp$ is the
distributive (Boolean) lattice of subsets of $\P$. If $\P=\PA$ is the
pure state space of a $C^*$-algebra $\Ac$ then ${\cal L}(\PA)$ may be
shown to be isomorphic (as an orthocomplemented lattice) to the
lattice of all projections in the von Neumann algebra $\pi_{\rm
ra}(\Ac)^-$.

For general compact convex sets it is not clear to what extent $\eb$
as a transition probability space equipped with the $w^*$-topology
determines $K$. If, however, $K=\SA$ is the state space of a unital
$C^*$-algebra $\A_{\C}$ (with self-adjoint part $\A$), then one {\em
can} reconstruct $\A$ as a $JB$-algebra, and hence the state space
$\SA$, from the pure state space $\PA$ as a \tps\ (with \tpies\ given
by (\ref{tpinca})), equipped with the $w^*$-uniformity (this is the
uniformity
\ci{Kel} $\cal U$   generated by sets of the form
$\{(\rh,\sg)\in\P\times\P|\,|\la
\rh-\sg,A\ra| <\varep\}$ for some 
$\varep>0$ and $A\in\A$; the physical interpretation of such
uniformities has been discussed by Haag, Kastler, and Ludwig, cf.\
\ci{Wer} for a very clear discussion.)

The essential step in this reconstruction is the following
reformulation of a result of Shultz
\ci{Shu} (whose formulation involved 
 $\pi_{\rm ra}(\Ac)^-$ rather than $\A(\PA)$) and Brown
\ci{Bro}: if $\A$ is the self-adjoint part of a unital $C^*$-algebra then
\be
\A = \A(\PA)\cap C_u(\PA), \ll{shultzeq}
\ee
where $C_u(\PA)$ is the space of uniformly continuous functions on
$\PA$, and, as before, $\A$ has been identified with a subspace of
$C(\PA)$ through the Gel'fand transform.  Note that to recover $\Ac$
as a $C^*$-algebra from the pure state space $\PA$, one in addition
needs an orientation of $\PA$, see \ci{AHS,Shu} and section 4.3.

For certain \ca s (called {\bf perfect}, cf.\ \ci{Shu,AS}) one can
replace $C_u(\PA)$ by $C(\PA)$ (with respect to the $w^*$-topology).
These include $\BH$ and $\KH$, for any Hilbert space $\H$.
\subsection{Poisson spaces with a transition probability}
Our goal, then, is to give axioms on a well-behaved \tps\ $\P$ which
enable one to construct, by a unified procedure, a $C^*$-algebra {\em
or} a Poisson algebra, which has $\P$ as its space of pure states, and
reproduces the original transition probabilities.  Moreover, even if
one is not interested in these axioms and takes quantum mechanics
(with superselection rules) at face value, the structure laid out in
this paper provides a transparent reformulation of quantum mechanics,
which may prove useful in the discussion of the classical limit
\ci{NPLcl}. 

 We first have to define a number of concepts, which play a
foundational role in both classical and quantum mechanics.  Apart from
transition probabilities, Poisson brackets play a central role in
dynamical theories.  Recall that a {\bf Poisson manifold} \ci{Wei,MR}
is a manifold $P$ with a Lie bracket $\{\,
,\,\}:\cin(P)\times\cin(P)\raw \cin(P)$, such that $\cin(P)$ equipped
with this Lie bracket, and pointwise multiplication as the Jordan
product $\circ$, is a Poisson algebra.  Symplectic manifolds are
special instances of Poisson manifolds; in the symplectic case the
Hamiltonian vector fields span the tangent space at every point of
$P$.  Recall from classical mechanics \ci{MR} that any $H\in\cin(P)$
defines a so-called {\bf Hamiltonian vector field} $X_H$ by
$X_Hf=\{H,f\}$; the flow of $X_H$ is called a {\bf Hamiltonian flow};
similarly, one speaks of a {\bf Hamiltonian curve}.

The most important result in the theory of Poisson manifolds states
that a Poisson manifold $P$ admits a decomposition into symplectic
leaves \ci{Wei,MR}. This means that there exists a family $S_{\al}$ of
symplectic manifolds, as well as continuous injections $\iota_{\al}:
S_{\al}\raw P$, such that $P=\cup_{\al}\iota_{\al}(S_{\al})$ (disjoint
union), and
\be
\{f,g\}(\iota_{\al}(\sg))=\{\iota_{\al}^*f,\iota_{\al}^*g\}_{\al}
(\sg),\ll{nieuw}
\ee
for all $\al$ and all $\sg\in S_{\al}$. Here $\{\, ,\,\}_{\al}$ is the
Poisson bracket associated to the symplectic structure on $S_{\al}$
\ci{MR}, and $(\iota_{\al}^*f)(\sg)=f(\iota_{\al}(\sg))$, etc.

We will need a generalization of the notion of a Poisson manifold,
which is inspired by the above decomposition.  \begin{definition} A
{\bf Poisson space} $P$ is a Hausdorff topological space together with
a linear subspace $\A\subset C(P)$ and a collection $S_{\al}$ of
symplectic manifolds, as well as continuous injections $\iota_{\al}:
S_{\al}\raw P$, such that:
\begin{itemize}
\item 
 $P=\cup_{\al}\iota_{\al}(S_{\al})$ (disjoint union);
\item
 $\A$ separates points;
\item
  $\A\subseteq \cin_L(P)$, where $\cin_L(P)$ consists of all $f\in
C(P)$ for which $\iota_{\al}^*f\in\cin(S_{\al})$ for each $\al$;
\item  $\A$ is closed under Poisson brackets. 
\end{itemize} \ll{defpoissonspace}
\end{definition}
The last requirement means, of course, that the Poisson bracket,
computed from the symplectic structure on the $S_{\al}$ and the above
decomposition of $P$ through (\ref{nieuw}), maps $\A\times\A$ into
$\A$.  In the context of Poisson spaces, each subspace
$\iota_{\al}(S_{\al})$ of $P$ is called a {\bf symplectic leaf} of
$P$. This terminology is sometimes applied to the $S_{\al}$ themselves
as well.

 In general, this decomposition falls under neither foliation theory
nor (Whitney) stratification theory (cf.\
\ci{SL} for this theory in a symplectic context).

If the ambient space $P$ carries additional structure, such as a
uniformity, or a smooth structure, one can refine the above definition
in the obvious way; such refinements will play an important role in
what follows.
\begin{definition}
 A {\bf uniform Poisson space} is a Poisson space $P$ in which the
topology is defined by a uniformity on $P$, and which satisfies
Definition \ref{defpoissonspace} with $C(P)$ replaced by
$C_u(P)$.\ll{defunpoissonspace}
\end{definition}
 Here $C_u(P)$ is the space of uniformly continuous functions on $P$;
it follows that elements of $\cin_L(P)$ are now required to be
uniformly continuous.

Similarly, a {\bf smooth Poisson space} is a Poisson space for which
$P$ is a manifold, and $C(P)$ in Definition \ref{defpoissonspace} is
replaced by $\cin(P)$. Hence $\cin_L(P)=\cin(P)$.  By the symplectic
decomposition theorem, a smooth Poisson space with $\A=\cin(P)$ is
nothing but a Poisson manifold.

In any case, $\cin_L(P)$ is the function space intrinsically related
to a (general, uniform, or smooth) Poisson space $P$.

The pure state space $\PA$ of a $C^*$-algebra $\Ac$ is a uniform
Poisson space in the following way.  We refer to (\ref{pssdec}) and
subsequent text.

Firstly, it follows directly from the definition of the
$w^*$-uniformity on $\PA$ that each $\hat{A}$, $A\in\A$, is in
$C_u(\PA)$; hence $\A\subset C_u(\PA)$, as required. As is well known,
a $C^*$-algebra separates the points of its pure state space (cf.\
\ci{KR1}).

Secondly, it is not difficult to show that the natural manifold
topology on a projective Hilbert space $\PH$ coincides with the
$w^*$-topology it inherits from the canonical inclusion $\PH\subset
{\cal S}(\BH)^*$. It follows that the inclusion map of any sector
$\PH_{\al}$ (equipped with the manifold topology) into $\P(\A)$ (with
the $w^*$-topology) is continuous.

Thirdly, there is a unique Poisson structure $\{\, ,\, \}$ on $\PA$
such that
\be
\{\hat{A},\hat{B}\}=i\widehat{(AB-BA)}.\ll{fust}
\ee
This Poisson bracket is defined by letting the sectors $\PH_{\al}$ of
$\PA$ coincide with its symplectic leaves, and making each $\PH_{\al}$
into a symplectic manifold by endowing it with the (suitably
normalized) Fubini-Study symplectic form
\ci{Str,Mar74,CLM,CMP1,CMP2,MR}.  The reason that this structure is
uniquely determined by (\ref{fust}) is that in an irreducible
representation $\pi(\Ac)$ on a \Hs\ $\H$ the collection of
differentials $\{d\widehat{\pi(A)},A\in\A\}$ is dense in the cotangent
space at each point of $\PH$.  Note that the precise choice of
$\H_{\al}$ in its unitary equivalence class does not affect the
definition of this Poisson structure, since it is invariant under
unitary transformations.  Since $\Ac$ is a $C^*$-algebra, $\A$ is
closed under the right-hand side of (\ref{fust}), and therefore under
the Poisson bracket on the left-hand side as well.

We now return to general Poisson spaces.\\ If $\P$ is simultaneously a
(general, uniform, or smooth) Poisson space and a \tps, two function
spaces are intrinsically associated with it: $\cin_L(\P)$ and $\AP$,
respectively. The space naturally tied with both structures in concert
is therefore
\be
\A_L(\P)=\AP\cap \cin_L(\P).\ll{defatildep}
\ee

Since elements of $\A_L(\P)$ are smooth on each symplectic leaf of
$\P$, they generate a well-defined Hamiltonian flow, which, of course,
stays inside a given leaf.
\begin{definition} 
 A (general, uniform, or smooth) Poisson space which is simultaneously
a \tps\ is called {\bf unitary} if the Hamiltonian flow on $\P$
defined by each element of $\A_L(\P)$ preserves the \tpies. That is,
if $\rh(t)$ and $\sg(t)$ are Hamiltonian curves (with respect to a
given $H\in \A_L(\P)$) through $\rh(0)=\rh$ and $\sg(0)=\sg$,
respectively, then
\be
p(\rh(t),\sg(t))=p(\rh,\sg)\ll{unitarityeq}
\ee
for each $t$ for which both flows are defined.
\ll{defunitarity}
\end{definition}

We now come to the central concept of this work.
\begin{definition} 
A {\bf (general, uniform, or smooth) Poisson space with a \tp} is a
set $\P$ which is a well-behaved
\tps\  and a unitary (general, uniform, or smooth) Poisson space, for which
$\A=\A_L(\P)$.
\ll{defupstp}
\end{definition}

This definition imposes two closely related compatibility conditions
between the Poisson structure and the \tpies: firstly, it makes a
definite choice for the space $\A$ appearing in the definition of a
Poisson space, and secondly it imposes the unitarity requirement.
 
  If $(\P,p)$ is a classical \tps\ (that is, $p$ is given by
(\ref{tpcm})), then any Poisson structure is unitary. This is, indeed,
the situation in classical mechanics, where $\P$ is the phase space of
a physical system.  The best-known example is, of course, $\P=\R^{2n}$
with canonical symplectic structure.

The pure state space $\PA$ of a $C^*$-algebra is a uniform Poisson
space with a \tp. Indeed, we infer from (\ref{oldapaispira}) that
$\AP\subset\cin_L(\PA)$, so that $\A_L(\PA)$ as defined in
(\ref{defatildep}) coincides with $\A$ as given in (\ref{shultzeq}).
Moreover, the flow of each $A\in\A$ on a given symplectic leaf (=
sector) $\PH_{\al}$ of $\PA$ is the projection of the flow
$\Ps(t)=\exp(-itA)\Ps$ on $\H_{\al}$.  Since $A$ is self-adjoint,
$\exp(-itA)$ is a unitary operator, and the \tpies\ (\ref{tpsforca})
are clearly invariant under such flows.  \section{Axioms for pure
state spaces}\eo As remarked above, a direct translation of the Axioms
AHS1-AHS5 for compact convex sets to axioms on their extreme
boundaries is difficult. Nevertheless, we can work with a set of
axioms on a set $\P$, some of which are similar to AHS1-AHS5. In
particular, AHS2 can be directly translated:
\begin{definition} 
A well-behaved \tps\ $\P$ said to have the {\bf two-sphere property}
if for any two points $\rh,\sg$ (with $\rh\neq \sg$) lying in the same
sector of $\P$, the space $\{\rh,\sg\}^{\perp\perp}$ is isomorphic as
a \tps\ to the two-sphere $S^2$, with \tpies\ given by
$p(z,w)=\half(1+\cos\th(z,w))$ (where $\th(z,w)$ is the angular
distance between $z$ and $w$, measured along a great
circle).\ll{deftwosphere} \end{definition} Here the orthoclosed space
$\{\rh,\sg\}^{\perp\perp}=\rh\vee\sg$ may be regarded as an element of
the lattice $\lp$.  If $\rh$ and $\sg$ lie in different sectors of
$\P$, then $\rh\vee\sg=\{\rh,\sg\}$; this follows from repeated
application of De Morgan's laws \ci{Kal1} and $\rh^{\perp\perp}=\rh$
(etc.).

To understand the nature of the two-sphere property, note that a
two-sphere $S^2$ with radius 1 may be regarded as the extreme boundary
of the unit ball $B^3\subset \R^3$, seen as a compact convex set.  As
we saw in section 1.2, $B^3\simeq {\cal S}(\M_2(\C))$.  Restricted to
the extreme boundary, the parametrization (\ref{parrho}) leads to a
bijection between $\P(\M_2(\C))\simeq\Bbb P\C^2$ and $S^2$. Under this
bijection the \tpies\ (\ref{tpsforca}) on $\Bbb P\C^2$ are mapped into
the ones stated in Definition \ref{deftwosphere}.

 In other words, the two-sphere property states that there exists a
fixed reference two-sphere $S_{\rm ref}^2\simeq \Bbb P\C^2$, equipped
with the standard \Hs\ \tpies\ $p=p_{\C^2}$ given by (\ref{tpsforca}),
and a collection of bijections $T_{\rh\vee\sg}: \rh\vee\sg \raw S_{\rm
ref}^2$, defined for each orthoclosed subspace of the type
$\rh\vee\sg\subset \P$ (where $\rh$ and $\sg\neq\rh$ lie in the same
sector of $\P$), such that
\be
p_{\C^2}(T_{\rh\vee\sg}(\rh'),T_{\rh\vee\sg}(\sg'))=p(\rh',\sg')\ll{Tiso}
\ee
 for all $\rh',\sg'\in
\rh\vee\sg$. 

Now consider the following axioms on a set $\P$:
\begin{trivlist}   
\item[] {\bf Axiom 1}: {\em $\P$ is a uniform Poisson space with a
 transition probability};
\item[] {\bf Axiom 2}: {\em $\P$ has the two-sphere property};
\item[] {\bf Axiom 3}: {\em the sectors of $\P$ as a \tps\ coincide 
with the symplectic leaves of
$\P$ as a Poisson space}; \item[] {\bf Axiom 4}: {\em the space $\A$
(defined through Axiom 1 by (\ref{defatildep})) is closed under the
Jordan product constructed from the \tpies};
\item[] {\bf Axiom 5}:  {\em  the pure state space
$\PA$ of $\A$ coincides with $\P$}.
\end{trivlist}

The meaning of Axiom 4 will become clear as soon as we have explained
how to construct a Jordan product on $\AP$, for certain \tps s $\P$.
This axiom turns $\A$ into a $JB$-algebra, which is contained in
$C(\P)$. Hence each element of $\P$ defines a pure state on $\A$ by
evaluation; Axiom 5 requires that all pure states of $\A$ be of this
form (note that, by Axiom 1, $\A$ already separates points).

 Axioms 2 and 4 are direct analogues of Axioms AHS2 and AHS4,
respectively (also cf.\ the end of section 4.2). The `bootstrap' Axiom
5 restricts the possible uniformities on $\P$; it is somewhat
analogous to Axiom AHS3.

In the previous section we have seen that the pure state space of a
unital $C^*$-algebra satisfies Axioms 1-5.

The remainder of this paper is devoted to the proof of the following
 
 {\bf Theorem }{\em If a set $\P$ satisfies Axioms 1-5 (with $\P$ as a
\tps\ containing no sector of dimension 3), then there exists a unital
\ca\ $\A_{\C}$, whose self-adjoint part is $\A$ (defined through Axiom
1). This $\Ac$ is unique up to isomorphism, and can be explicitly
reconstructed from $\P$, such that
\begin{enumerate}
\item $\P=\PA$ (i.e., $\P$ is the pure state space of $\A$);
\item the transition probabilities (\ref{mtp}) coincide
 with those initially given on $\P$;
\item the Poisson structure on each symplectic leaf of $\P$ is
 proportional to the Poisson structure
imposed on the given leaf by (\ref{fust});
\item the $w^*$-uniformity on $\PA$ defined by $\A$ is contained 
in the initial uniformity on $\P$;
\item the $C^*$-norm on $\A\subset \Ac$ is equal to the sup-norm 
inherited from the inclusion
$\A\subset \ell^{\infty}(\P)$.
\end{enumerate}}
 
The unfortunate restriction to \tps s without 3-dimensional sectors
(where the notion of dimension is as defined after (\ref{tp4}), i.e.,
the cardinality of a basis of $\P$ as a \tps) follows from our method
of proof, which uses the von Neumann co-ordinatization theorem for
Hilbert lattices \ci{FH1,Var,Kal2}. In view of the parallel between
our axioms and those in \ci{AHS}, however, we are confident that the
theorem holds without this restriction.  To make progress in this
direction one has to either follow our line of proof and exclude the
possibility of non-Desarguesian projective geometries (cf.\
\ci{FH1,FH2} in the present context),
 or abandon the use of Hilbert lattices and develop a spectral theory
of well-behaved \tps s, analogous to the spectral theory of compact
convex sets of Alfsen and Shultz \ci{AS1,AS2}. Despite considerable
efforts in both directions the author has failed to remove the
restriction.

The theorem lays out a possible mathematical structure of quantum
mechanics with superselection rules. Like all other attempts to do so
(cf.\
\ci{vN1,vN2,Seg,Lud,BC}), the axioms appear to be contingent.  
This is particularly true of Axiom AHS2 and of our Axiom 2, which lie
at the heart of quantum mechanics. One advantage of the axiom schemes
in \ci{AHS} and the present paper is that they identify the incidental
nature of quantum mechanics so clearly.

If $\P$ is merely assumed to be a Poisson space with a \tp\ (i.e., no
uniformity is present), then the above still holds, with the obvious
modifications. In that way, however, only perfect \ca s
\ci{Shu,AS} can be reconstructed (cf.\ section 1.4).
 \section{From transition probabilities to $C^*$-algebras} \eo The
proof of the theorem above essentially consists of the construction of
a \ca\ $\Ac$ from the given set $\P$.  In summary, we can say that in
passing from pure states to algebras of observables one has the
following correspondences.
\begin{center}
\begin{tabular}{|lr|}\hline
{\em Pure state space} & {\em Algebra of observables}\\
\mbox{} & \mbox{} \\
transition probabilities & Jordan product\\ Poisson structure &
Poisson bracket\\ unitarity & Leibniz rule\\ \hline
\end{tabular}
\end{center}

To avoid unnecessary interruptions of the argument, some of the more
technical arguments are delegated to Chapter 4.
\subsection{Identification of $\P$ as a \tps} 
 This identification follows from Axiom 1 (of which only the part
stating that $\P$ be a well-behaved \tps\ is needed) and Axiom 2, as a
consequence of the following result.
\begin{proposition} 
Let a well-behaved \tps\ $\P$ (with associated lattice $\LP$) have the
two-sphere property.  If $\P$ has no sector of dimension 3, then
$\P\simeq \cup_{\al}\PH_{\al}$ as a \tps\ (for some family
$\{\H_{\al}\}$ of complex \Hs s), where each sector $\PH_{\al}$ is
equipped with the \tpies\ (\ref{tpsforca}).
\ll{chofp}
\end{proposition}
 This statement is not necessarily false when $\P$ does have sectors
of dimension 3 (in fact, we believe it to be always true in that case
as well); unfortunately our proof does not work in that special
dimension.

In any case, it is sufficient to prove the theorem for each sector
separately, so we may assume that $\P$ is irreducible (as a \tps).
Even so, the proof is quite involved, and will be given in section
4.1.
\subsection{Spectral theorem} 
For each orthoclosed subset $Q$ of a well-behaved \tps\ $\P$, define a
function $p_Q$ on $\P$ by
\be
p_Q=\sum_{i=1}^{\dim(Q)} p_{e_i} \ll{defpqeq};
\ee
here is $\{e_i\}$ is a basis of $Q$; it is easily seen that $p_Q$ is
independent of the choice of this basis (cf.\ \ci{Zab}).
\begin{definition}
Let $\P$ be a well-behaved \tps. A {\bf spectral resolution} of an
element $f\in\ell^{\infty}(\P)$ is an expansion (in the topology of
pointwise convergence)
\be
f=\sum_j \lm_j p_{Q_j},\ll{defspectral}
\ee
where $\lm_j\in \R$, and $\{Q_j\}$ is an orthogonal family of
orthoclosed subsets of $\P$ (cf.\ (\ref{defpqeq})) for which $\sum_j
p_{Q_j}$ equals the unit function on $\P$.
\ll{defspectralthm}
\end{definition}
\begin{proposition} If $\P=\cup_{\al}\PH_{\al}$ (with \tpies\
(\ref{tpsforca})) then any $f\in\A_{00}(\P)$ has a unique spectral
resolution.
\ll{spectraltheorem}
\end{proposition}

By the previous section this applies, in particular, to a \tps\ $\P$
satisfying Axioms 1 and 2.

{\em Proof}.  Firstly, the case of reducible $\P$ may be reduced to
the irreducible one by grouping the $\rh_i$ in $f = \sum_{i=1}^N c_i
p_{\rh_i}$ into mutually orthogonal groups, with the property that
$(\cup \rh)^{\perp\perp}$ is irreducible if the union is over all
$\rh_i$ in a given group.  Thus we henceforth assume that $\P$ is
irreducible, hence of the form $\P=\PH$ with the \tpies\
(\ref{tpsforca}).

If $\P$ is finite-dimensional the proposition is simply a restatement
of the spectral theorem for Hermitian matrices.  In the general case,
let $f $ be as above, and $Q:=\{ \rh_1,\ldots ,\rh_N\}^{\perp\perp}$.
If $\sg\in Q$ then $f (\sg)=\sum_j \lm_j p_{Q_j}(\sg)$ for some
$\lm_j$ and mutually orthogonal $Q_j\subset Q$, as in the previous
paragraph. If $\sg\in Q^{\perp}$ this equation trivially holds, as
both sides vanish.

Let us assume, therefore, that $\sg$ lies neither in $Q$ nor in
$Q^{\perp}$.  Define $\phv_Q(\sg)$ by the following procedure: lift
$\sg$ to a unit vector $\Sigma$ in $\H$, project $\Sg$ onto the
subspace defined by $Q$, normalize the resulting vector to unity, and
project back to $\PH$ (this is a Sasaki projection in the sense of
lattice theory \ci{BC,Kal1}).  In the Hilbert space case relevant to
us, the transition probabilities satisfy
\be
p(\sg,\rh)=p(\sg,\phv_Q(\sg))p(\phv_Q(\sg),\rh)
\ee
 for $\rh\in Q$ and $\sg\notin Q^{\perp}$.  We now compute $f (\sg)$
by using this equation, followed by the use of the spectral theorem in
$Q$, and subsequently we recycle the same equation in the opposite
direction. This calculation establishes the proposition for $\sg\notin
Q^{\perp}$. \enp

If $\P$ is a classical \tps\ (see (\ref{tpcm})) then a spectral
theorem obviously holds as well; it simply states that a function $f$
with finite support $\{\sg_i\}$ is given by $f=\sum_i
f(\sg_i)p_{\sg_i}$.  \subsection{Jordan structure\ll{Jost}}
\begin{proposition} 
If $\P=\cup_{\al}\PH_{\al}$ (with \tpies\ (\ref{tpsforca})), $f=\sum_j
\lm_j p_{Q_j}$ is the spectral resolution of $f\in \A_{00}(\P)$, and
$f^2$ is defined by $f^2=\sum_j \lm_j^2 p_{Q_j}$, then the product
$\circ$ defined by
\be
f\circ g=\quar ((f+g)^2-(f-g)^2) \ll{cofjp}
\ee
turns $\A_{00}(\P)$ into a Jordan algebra. Moreover, this Jordan
product $\circ$ can be extended to $\AOP$ by (norm-) continuity, which
thereby becomes a \jba\ (with the sup-norm inherited from
$\ell^{\infty}(\P)$). Finally, the bidual $\AP$ is turned into a \jba\
by extending $\circ$ by $w^*$-continuity.
\ll{fromtptojp}
\end{proposition} 

The bilinearity of (\ref{cofjp}) is not obvious, and would not
necessarily hold for arbitrary well-behaved \tps s in which a spectral
theorem (in the sense of Proposition \ref{spectraltheorem}) is valid.
In the present case, it follows, as a point of principle, from the
explicit form of the \tpies\ in $\PH$. The quickest way to establish
bilinearity, of course, is to look at a function $p_Q$ (where $Q$ lies
in a sector $\PH$ of $\P$) as the Gel'fand transform of a projection
operator on $\H$.

Given bilinearity, the claims of the proposition follow from the
literature.  The extension to $\AOP$ by continuity, turning it into a
$JB$-algebra, is in
\ci[Thm.\ 12.12]{AS1} or \ci[Prop.\ 6.11]{AS3}.
For the the extension to $\AP$ see section 3 of \ci{ASS} and section 2
and Prop.\ 6.13 of \ci{AS3}. (There is a spectral theorem in $\AP$,
which is a so-called $JBW$-algebra, as well, cf.\ \ci{AS1,AS2,ASS},
but we will not need this.)

The norm in $\AP$ is the sup-norm inherited from $\ell^{\infty}(\P)$
as well; this establishes item 5 of the Theorem.

If $\P$ is classical, $\AP=\ell^{\infty}(\P)$, and the Jordan product
constructed above is given by pointwise multiplication. This explains
why the latter is used in classical mechanics.  \subsection{Explicit
description of $\AP$\ll{aopdd}}
\begin{proposition} Let $\P=\cup_{\al}\PH_{\al}$ (with \tpies\
(\ref{tpsforca})), and regard self-adjoint elements
$A=\oplus_{\al}A_{\al}$ of the von Neumann algebra
$\M_{\C}=\oplus_{\al}{\frak B}(\H_{\al})$ as functions $\hat{A}$ on
$\P$ in the obvious way: if $\rh\in \PH_{\al}$ then
$\hat{A}(\rh)=\rh(A_{\al})$.  Denote the subspace of
$\ell^{\infty}(\P)$ consisting of all such $\hat{A}$, $A\in \M$, by
$\hat{\M}$.  Then
\ll{propapa}
\end{proposition}
\be
\A(\P)=\hat{\M}. \ll{apaispira}
\ee
Note that the identification of $A\in\M$ with
$\hat{A}\in\ell^{\infty}(\P)$ is norm-preserving relative to the
operator norm and the sup-norm, respectively.  Also, it is clear that
this proposition proves (\ref{oldapaispira}).

{\em Proof.} Inspired by \ci{ACLM,CMP1}, we define a (locally
non-trivial) fiber bundle ${\cal B}(\P)$, whose base space $B$ is the
space of sectors, equipped with the discrete topology, and whose fiber
above a given base point $\al$ is ${\frak B}(\H_{\al})_{\rm sa}$; here
$\H_{\al}$ is such that the sector $\al$ is ${\Bbb P}\H_{\al}$.
Moreover, $\P$ itself may be seen as a fiber bundle over the same base
space; now the fiber above $\al$ is $\PH_{\al}$. We will denote the
projection of the latter bundle by $pr$.  A cross-section $s$ of
${\cal B}(\P)$ then defines a function $\hat{s}$ on $\P$ by
$\hat{s}(\rh)= [s(pr(\rh))](\rh)$.  The correspondence
$s\leftrightarrow \hat{s}$ is isometric if we define the norm of a
cross-section of ${\cal B}(\P)$ by $\n s\n=\sup_{\al\in B} \n
s(\al)\n$ (where the right-hand side of course contains the operator
norm in ${\frak B}(\H_{\al})$), and the norm of $\hat{s}$ as the
sup-norm in $\ell^{\infty}(\P)$.

 It follows directly from its definition that the space $\A_{00}(\P)$
consists of sections $s$ of ${\cal B}(\P)$ with finite support, and
such that $s(\al)$ has finite rank for each $\al$. Its closure $\AOP$
contains all sections such that the function $\al\raw \n s(\al)\n$
vanishes at infinity, and $s(\al)$ is a compact operator.  It follows
from elementary operator algebra theory that the dual $\AOP^*$ may be
realized as the space of sections for which $s(\al)$ is of trace-class
and $\al\raw\n s(\al)\n_1$ (the norm here being the trace-norm) is in
$\ell^1(B)$. The bidual $\AP$ then consists of all sections of ${\cal
B}(\P)$ for which $\al\raw \n s(\al)\n$ is in $\ell^{\infty}(B)$ (here
the crucial point is that ${\frak K}(\H)^{**}={\frak B}(\H)$). Eq.\
(\ref{apaispira}) is then obvious.  \enp

  For later use, we note that $\AOP$ and even $\A_{00}(\P)$ are dense
in $\AP$ in the topology of pointwise convergence.  This is because
firstly ${\frak K}(\H)$ is dense in ${\frak B}(\H)$ in the weak
operator topology \ci{KR1} (as is the set of operators of finite
rank), hence certainly in the coarser topology of pointwise
convergence on $\P$, and secondly the topology of pointwise
convergence on $\ell^{\infty}(B)$ is contained in the $w^*$-topology
($\ell^{\infty}(B)$ being the dual of $\ell^1(B)$, which in turn is
the dual of $\ell_0(B)$); recall that any (pre-) Banach space is
$w^*$-dense in its double dual (e.g., \ci{KR1}).

Under the correspondence $\AP=\hat{\M}\leftrightarrow \M$ the Jordan
product constructed in the previous section is then simply given by
the anti-commutator of operators in $\M$.
\subsection{Algebra of observables} 
By Axiom 1, the space of observables $\A$ is defined by
(\ref{defatildep}).  We now use Axiom 3, which implies that each
symplectic leaf of $\P$ is a projective
\Hs\ $\PH_{\al}$. 
For the moment we assume that each leaf $\PH_{\al}$ has a manifold structure
relative to which all functions $\hat{A}$, where $A\in {\frak
B}(\H_{\al})_{\rm sa}$, are smooth (such as its usual manifold
structure). Then $\AP\cap C_u(\P)\subset \cin_L(\P)$ by the explicit
description of $\AP$ just obtained.  It then follows from
(\ref{defatildep}) that
\be
\A=\AP\cap C_u(\P). \ll{shultzbis}
\ee

It is easily shown that $\A$ is closed (in the sup-norm). This follows
from the fact that $\AP$ is closed, plus the observation that the
subspace of functions in $\ell^{\infty}(\P)$ which are uniformly
continuous with respect to any uniformity on $\P$, is closed; this
generalizes the well-known fact that the subspace of continuous
functions relative to any topology on $\P$ is sup-norm closed (the
proof of this observation proceeds by the same
$\varepsilon/3$-argument.)

 Note that $\AOP$ is not necessarily a subspace of $\A$; it never is
if the $C^*$-algebra $\A_{\C}$ to be constructed in what follows is
antiliminal \ci{Dix}.
 
We can construct a Jordan product in $\A$ by the procedure in section
\ref{Jost}. By Proposition \ref{fromtptojp} and Axiom 4, this turns
$\A$ into a \jba.  At this stage we can already construct the pure
state space $\PA$; the first claim of the Theorem then holds by Axiom
5.

 We may regard the restriction of $\A$ to a given sector $\PH_{\al}$
as the Gel'fand transform of a Jordan subalgebra of ${\frak
B}(\H_{\al})_{\rm sa}$.  This subalgebra must be weakly dense in
${\frak B}(\H_{\al})_{\rm sa}$, for otherwise Axiom 5 cannot hold.

Let us now assume that some $\PH_{\al}$ have an exotic manifold
structure such that $\AP\cap C_u(\P)$ is not contained in
$\cin_L(\P)$, so that $\A\subset \AP\cap C_u(\P)$ is a proper
inclusion (rather than the equality (\ref{shultzbis})). It follows
from Axiom 5 that the statement in the previous paragraph must still
hold.  This weak density suffices for the results in sections 3.7 and
3.8 to hold, and we can construct a
\ca\  $\Ac$ with pure state space $\P$.
 The proper inclusion above would then contradict
(\ref{shultzeq}). Hence such exotic manifold structures are excluded
by the axioms.  \subsection{Unitarity, Leibniz rule, and Jordan
homomorphisms} It is instructive to discuss a slightly more general
context than is strictly necessary for our purposes.
\begin{proposition} Let $\P$ be a Poisson space with a \tp\ in which
every $f\in \A_{00}(\P)$ has a unique spectral resolution (in the
sense of Definition \ref{defspectralthm}).  Assume that for each $H\in
\A_L(\P)$ (cf.\ (\ref{defatildep})) the map $f\raw \{H,f\}$ is bounded
on $\A_L(\P)\subset\ell^{\infty}(\P)$ (with sup-norm). If a Jordan
product $\circ$ is defined on $\A_L(\P)$ through the \tpies, in the
manner of Proposition \ref{fromtptojp}, then $\circ$ and the Poisson
bracket satisfy the Leibniz rule (\ref{leib}).
\ll{unitaritytoleibniz}  
\end{proposition}

The boundedness assumption holds in the case at hand (cf.\ the next
section); it is mainly made to simplify the proof. The proposition
evidently holds when $\A_L(\P)$ is a Poisson algebra, for which the
assumption is violated.

{\em Proof.} Writing $\dl_H(f)$ for $\{H,f\}$, the boundedness of
$\dl_H$ implies that the series $\al_t(f)=\sum_{n=0}^{\infty} t^n
\dl_H^n(f)/n!$ converges uniformly, and defines a uniformly continuous
one-parameter group of maps on $\A_L(\P)$ (cf.\ \ci{BR1}).  On the
other hand, if $\sg(t)$ is the Hamiltonian flow of $H$ on $\P$ (with
$\sg(0)=\sg$), then $\al_t$ as defined by $\al_t(f):\sg\raw f(\sg(t))$
must coincide with the definition above, for they each satisfy the
same differential equation with the same initial condition. In
particular, the flow in question must be complete.  Moreover, it
follows that the Leibniz rule (yet to be established) is equivalent to
the property that $\al_t$ is a Jordan morphism for each $t$; this, in
turn, can be rephrased by saying that $\al_t(f^2)=\al_t(f)^2$ for all
$f\in \A_L(\P)$.

 Let $f\in\A_{00}(\P)\cap \A_L(\P)$, so that $f=\sum_k \lm_k p_{e_k}$,
where all $e_k $ are orthogonal (cf.\ section 3.2). Unitarity implies
firstly that $\al_t(f)=\sum_k \lm_k p_{e_k(-t)}$, and secondly that
the $e_k(-t)$ are orthogonal. Hence $\al_t(f)$ is given in its
spectral resolution, so that $\al_t(f)^2=\sum_k \lm_k^2 p_{e_k(-t)}$.
Repeating the first use of unitarity, we find that this equals
$\al_t(f^2)$. Hence the property holds on $\A_{00}(\P)$.

 Now $\A_{00}(\P)$ is dense in $\AP$ in the topology of pointwise
convergence in $\ell^{\infty}(\P)$.  But $f_{\lm}\raw f$ pointwise
clearly implies $\al_t(f_{\lm})\raw \al_t(f)$ pointwise.  This, plus
the $w^*$-continuity of the Jordan product \ci{ASS} proves the desired
result.
\subsection{Poisson structure} 
  Item 3 of the Theorem follows from Axiom 3, the penultimate
paragraph of section 3.5, and the following
\begin{proposition}
Let $\PH$, equipped with the \tpies\ (\ref{tpsforca}), be a unitary
Poisson space for which the Poisson structure is symplectic, and for
which $\A$ is the Gel'fand transform of a weakly dense subspace of
${\frak B}(\H_{\al})_{\rm sa}$.

 Then the Poisson structure is determined up to a multiplicative
constant, and given by (\ref{fust}) times some $\hbar^{-1}\in\R$.
\end{proposition}
 
{\em Proof}.  Axiom 3 implies that each sector $\Bbb P\H$ (for some
$\H$) is a symplectic space.  Unitarity (in our sense) and Wigner's
theorem (cf.\
\ci{Var,BC,Shu} for the latter) imply that each $\hat{A}\in\A$
   generates a Hamiltonian flow on $\Bbb
P\H$ which is the projection of a unitary flow on $\H$. Therefore,
$\{\hat{A},\hat{B}\}(\ps)=\frac{d}{dt}\hat{B}(\exp(it\hat{C}(A))\ps)_{t=0}$
for some self-adjoint operator $C$, depending on $A$ (here
$\exp(it\hat{C}(A))\ps$ is by definition the projection of
$\exp(itC(A))\Ps$ to $\PH$, where $\Ps$ is some unit vector in $\H$
which projects to $\ps\in\PH$).  The right-hand side equals
$i\widehat{(CB-BC)}(\ps)$. Anti-symmetry of the left-hand side implies
that $C=\hbar^{-1} A$ for some $\hbar^{-1}\in\R$. By the weak density
assumption, the collection of all differentials $d\hat{A}$ spans the
fiber of the cotangent bundle at each point of ${\Bbb P}\H$. Thus the
Poisson structure is completely determined. \enp

This shows that the symplectic structure on each leaf is
$\hbar\om_{FS}$, where $\om_{FS}$ is the Fubini-Study structure
\ci{Str,Mar74,CLM,CMP1,CMP2,MR}.  (A closely related fact is that the
K\"{a}hler metric associated to $\om_{FS}$ is determined, up to a
multiplicative constant, by its invariance under the induced action of
all unitary operators on $\H$, cf.\ \ci{ACLM,MR}.)  The multiplicative
constant is Planck's constant $\hbar$, which, as we see, may depend on
the sector.  To satisfy Axiom 4, $\hbar^{-1}$ must be nonzero in every
sector whose dimension is greater than 1. In one-dimensional sectors
the Poisson bracket identically vanishes, so that the value of $\hbar$
is irrelevant.

The Poisson structure on $\P$ is determined by the collection of
symplectic structures on the sectors of $\P$, for the Poisson bracket
$\{f,g\}(\rh)$ is determined by the restrictions of $f$ and $g$ to the
leaf through $\rh$; cf.\ (\ref{nieuw}).

 The choice (\ref{fust}) for the Poisson bracket on $\A$ corresponds
to taking $\hbar$ a sector-independent constant (put equal to 1).  In
general, we may regard $\hbar$ as a function on $\PA$, which is
constant on each sector.  If $\hat{A}$ denotes an element of $\A$, the
restriction of $\hat{A}$ to a sector $\PH_{\al}$ corresponds to an
operator $A_{\al}\in {\frak B}(\H_{\al})_{\rm sa}$ (cf.\ section 3.5).
The sector in which $\rh\in\PA$ lies is called $\al(\rh)$.  With this
notation, and denoting $AB-BA$ by $[[A,B]]$ (recall that $[A,B]$
denotes the Lie bracket in a Jordan-Lie algebra) the Poisson bracket
on $\A$ is then given by
\be
\{\hat{A},\hat{B}\}(\rh)=\frac{i}{\hbar(\rh)}
\widehat{[[A_{\al(\rh)}, B_{\al(\rh)} ]] }(\rh).\ll{hbarvaries} 
\ee

The sector-dependence of $\hbar$ cannot be completely arbitrary,
however; Axiom 1 implies that $\hbar$ must be a uniformly continuous
function on $\P$.  For suppose $\hbar$ is not uniformly continuous. We
then take $\hat{A},\, \hat{B}\in\A$ in such a way that $A_{\al}$ and
$B_{\al}$ are independent of $\al$ in a neighbourhood of a point $\sg$
of discontinuity of $\hbar$, with $[[A_{\al(\sg)},B_{\al(\sg)} ]]\neq
0$.  Then the real-valued function on $\PA$ defined by $\rh\raw
\hbar(\rh)\{\hat{A},\hat{B}\}(\rh)$ is certainly uniformly continuous
near $\sg$, since its value at $\rh$ is equal to
$i\widehat{[[A_{\al(\rh)},B_{\al(\rh)} ]]}(\rh)$.  But, by assumption,
$\{\hat{A},\hat{B}\}$ is uniformly continuous as well.  Because of the
factor $\hbar$, the product $\hbar \{\hat{A},\hat{B}\}$ cannot be
uniformly continuous. This leads to a contradiction.
\subsection{$C^*$-structure}
 We now turn $\A$ into a Jordan-Lie algebra, and thence into the
self-adjoint part of a $C^*$-algebra $\A_{\C}$ (cf.\ section 1.1).
 
On each leaf, the associator equation (\ref{ass}) is identically
satisfied by the Poisson bracket (\ref{hbarvaries}). However, the
`constant' $k\equiv
\hbar^2/4$ may depend on the leaf.
  Therefore, we have to rescale the Poisson bracket so as to undo
its $\hbar$-dependence. From (\ref{hbarvaries}) this is obviously
accomplished by putting $[f,g](\rh)=\hbar(\rh)\{f,g\}(\rh)$.  With the
Jordan product $\circ$ defined in section 3.3, eq.\ (\ref{ass}) is now
satisfied.  Hence we define a product $\cdot: \A\times\A\raw \A_{\C}$
by
\be
f\cdot g=f\circ g-\half i [f,g],
\ll{cstprod}
\ee
and extend this to $\Ac\times \Ac$ by complex linearity.

As explained in section 1.1, this product is associative. Indeed, in
the notation introduced in the previous section one simply has
\be
\hat{A}\cdot \hat{B}(\rh) = \widehat{A_{\al(\rh)} B_{\al(\rh)}} (\rh),
\ee
where the multiplication on the right-hand side is in ${\frak
B}(\H_{\al(\rh)})$.
 
By Axiom 1 (in particular, closure of $\A$ under the Poisson bracket),
Axiom 4, and the uniform continuity of $\hbar(\cdot)$, $\A_{\C}$ is
closed under this associative product.

Let $\A$ be a $JB$-algebra, and $\A_{\C}=\A\oplus i\A$ its
complexification. As shown in \ci{Wri}, one may construct a norm on
$\A_{\C}$, which turns it into a so-called $JB^*$-algebra \ci{HOS};
the involution is the natural one, i.e., $(f+ig)^*=f-ig$ for
$f,g\in\A$.  Now given a $JB^*$-algebra $\A_{\C}$ whose Jordan product
$\circ$ is the anti-commutator of some associative product $\cdot$, it
is shown in \ci{Rod} that $(\A_{\C},\cdot)$ is a $C^*$-algebra iff
$(\A_{\C},\circ)$ is $JB^*$-algebra.

Hence one can find a norm on $\A_{\C}$ (whose restriction to its
self-adjoint part $\A$, realized as in (\ref{shultzeq}), is the
sup-norm) such that it becomes a $C^*$-algebra equipped with the
associative product (\ref{cstprod}). Since the unit function evidently
lies $\AP$ (cf.\ (\ref{apaispira})) as well as in $C_u(\P)$, it lies
in in $\A$ (cf.\ (\ref{shultzbis})).  In conclusion, the unital \ca\
mentioned in the Theorem has been constructed.

An alternative argument showing that $\A$ is closed under the
commutator (Poisson bracket) is to combine the results of section 4.3
below and \ci[\S 7]{AHS}. This avoids the rescaling of the Poisson
bracket by $\hbar(\cdot)$, but relies on the deep analysis of
\ci{AHS}.
 
It is also possible to have $+$ instead of $-$ in (\ref{cstprod}).
This choice produces a $C^*$-algebra $\Ac^{(+)}$ which is canonically
anti-isomorphic to $\Ac\equiv \Ac^{(-)}$.  Moreover, in some cases
$\Ac^{(+)}$ is isomorphic to $\Ac^{(-)}$ in a non-canonical fashion.
Choose a faithful representation $\pi(\Ac)$ on some Hilbert space
$\H$, and choose a basis $\{{\bf e}_i\}$ in $\H$. Then define an
anti-linear map $J:\H\raw\H$ by $J\sum_i c_i {\bf e}_i=\sum_i
\ovl{c}_i {\bf e}_i$, and subsequently a linear map $j$ on $\pi(\Ac)$
by $j(A)=J\pi(A)^*J$. If $j$ maps $\pi(\Ac)$ into itself, it defines
an isomorphism between $\Ac^{(-)}$ and $\Ac^{(+)}$.

In \ci{AHS} (or \ci{Shu}) this sign change would correspond to
reversing the orientation of $K$ (or $\P$).
\subsection{Transition
probabilities and uniform structure} Recall Mielnik's definition
(\ref{mtp}) of the transition probability in the extreme boundary
$\partial_e K$ of a compact convex set \ci{Mie2}.

By Axiom 5, the extreme boundary of the state space $K=\SA$ of $\A$ is
$\P$.  Hence $\P$ acquires transition probabilities by (\ref{mtp}),
which are to be compared with those originally defined on it. In
section 4.2 we show that these transition probabilities coincide, and
this proves item 2 of the Theorem.
 
It is immediate from the previous paragraph that $\A(\PA)=\AP$.  The
$w^*$-uniformity appearing in (\ref{shultzeq}) is the weakest
uniformity relative to which all elements of $\A$ are uniformly
continuous.  It then follows from (\ref{shultzeq}) and
(\ref{shultzbis}) (in which the uniformity is the initially given one)
that the initial uniformity on $\P$ must contain the $w^*$-uniformity
it acquires as the space of pure states of $\A_{\C}$. This proves item
4.

This completes our construction, as well as the proof of the Theorem.
\enp \section{Proofs} \eo \subsection{Proof of Proposition
\protect\ref{chofp} } The strategy of the proof is to characterize the
lattice $\lp$ (cf.\ section 1.3), and then use the so-called
co-ordinatization theorem in lattice theory \ci{BC,Kal2} to show that
$\LP$ is isomorphic to the lattice $\L(\H)$ of closed subspaces of
some complex Hilbert space $\H$ (see \ci{Var,BC,Kal1,Kal2} for
extensive information on this lattice; an equivalent description is in
terms of the projections in the von Neumann algebra $\BH$).
 
It is known that $\lp$ is complete, atomic, and orthomodular
\ci{Zab,Bel,BC} if $\P$ is a well-behaved \tps; hence it is also
atomistic
\ci{BC,Kal1}. Using the connection between the
 center of an orthomodular lattice and its reducibility
\ci{Kal1}, it is routine to show that
 the irreducibility of $\P$ as a \tps\ (which we assume for the
purpose of this proof) is equivalent to the irreducibility of $\lp$ as
a lattice.  Hence $\lp$ is also irreducible.

{\bf Lemma 1} {\em $\lp$ has the covering property (i.e., satisfies
the exchange axiom).}\\ See \ci{BC,Kal1,Kal2} for the relevant
definitions and context.

{\em Proof.} Consistent with previous notation, we denote atoms of
$\LP$ (hence points of $\P$) by $\rh,\sg$, and arbitrary elements by
$Q,Q_i,R,S$.

Let $n=\dim(Q)$ (as a \tps); for the moment we assume $n<\infty$. We
will first use induction to prove that if $\rh\notin Q$, the element
$(\rh\vee Q)\wed Q^{\perp}$ is an atom.

To start, note that if $Q_1\leq Q_2$ for orthoclosed $Q_1, Q_2$ of the
same finite dimension, then $Q_1=Q_2$.  For an orthoclosed set in $\P$
is determined by a basis of it, which in turn determines its
dimension. This implies that $\dim(\rh\vee Q)>\dim(Q)$ if $\rh\notin
Q$ (take $Q_1=Q$ and $Q_2=\rh\vee Q$).  Accordingly, it must be that
$(\rh\vee Q)\wed Q^{\perp}>\emptyset $, for equality would imply that
$\dim(\rh\vee Q)=\dim(Q)$.

For $n=1$, $Q$ is an atom. By assumption, $\rh\vee Q$ is $S^2$, hence
$(\rh\vee Q)\wedge Q^{\perp}$ is the anti-podal point to $Q$ in
$\rh\vee Q$, which is an atom, as desired.  Now assume $n>1$. Choose a
basis $\{e_i\}_{i=1,\dots,\dim(Q)}$ of $Q$; then $Q=\vee_{i=1}^n e_i$.
Put $R=\vee_{i=1}^{n-1} e_i$; then $R<Q$ hence $Q^{\perp}< R^{\perp}$,
so that $(\rh\vee Q)\wed Q^{\perp} \leq (\rh\vee Q)\wed R^{\perp}$.
The assumption $(\rh\vee Q)\wed Q^{\perp} = (\rh\vee Q)\wed R^{\perp}$
is equivalent, on use of $Q=R\vee e_n$, De Morgan's laws \ci{Kal1},
and the associativity of $\wed$, to $((\rh\vee Q)\wed R^{\perp})\wed
e_n^{\perp}=(\rh\vee Q)\wed R^{\perp}$, which implies that $(\rh\vee
Q)\wed R^{\perp}\leq e_n^{\perp}$. This is not possible, since the
left-hand side contains $e_n$. Hence \be
\emptyset  < (\rh\vee Q)\wed Q^{\perp} < (\rh\vee Q)\wed R^{\perp}. \ll{l11}
\ee
It follows from the orthomodularity of $\lp$ that if $R\leq S$ and
$R\leq Q$, then
\be
 (S\vee Q)\wed R^{\perp}=(S\wed R^{\perp})\vee (Q\wed R^{\perp}).
\ll{l12}
\ee
Since $R<Q$ and $R\leq \rh\vee R$, one has $\rh\vee Q=(\rh\vee R)\vee
Q$. Now use (\ref{l12}) with $S=\rh\vee R$ to find $$ (\rh\vee Q)\wed
R^{\perp}=((\rh\vee R)\vee Q)\wed R^{\perp}=((\rh\vee R)\wed
R^{\perp})\vee (Q\wed R^{\perp}).  $$ By the induction hypothesis
$(\rh\vee R)\wed R^{\perp}$ is an atom (call it $\sg$), so the
right-hand side equals $\sg\vee e_n$.  The equality $\sg=e_n$ would
imply that $\rh\in Q$, hence $\sg\neq e_n$. But then (\ref{l11}) and
the $S^2$-assumption imply $0< \dim((\rh\vee Q)\wed Q^{\perp})<2$, so
that $(\rh\vee Q)\wed Q^{\perp}$ must indeed be an atom.

It follows that $\dim(\rh\vee Q)=\dim(Q)+1$.  Hence any $S\subset\P$
satisfying $Q\leq S\leq \rh\vee Q$ must have $\dim(S)$ equal to
$\dim(Q)$ or to $\dim(Q)+1$. In the former case, it must be that $S=Q$
by the dimension argument earlier. Similarly, in the latter case the
only possibility is $S=\rh\vee Q$. All in all, we have proved the
covering property for finite-dimensional sublattices.

A complicated technical argument involving the dimension theory of
lattices then shows that the covering property holds for all
$x\in\LP$; see \S 13 in \ci{Kal1} and \S 8 in \ci{Kal2}. \enp
 
 We have, therefore, shown that $\lp$ is a complete atomistic
irreducible orthomodular lattice with the covering property.  If $\lp$
is in addition infinite-dimensional, one speaks of a Hilbert lattice
(recently, there has been a major breakthrough in the theory of such
lattices \ci{Sol,Hol}, but since the infinite-dimensionality is used
explicitly in this work we derive no direct benefit from this).  In
any case, we are in a position to apply the standard co-ordinatization
theorem of lattice theory; see \ci{FH1,Var,BC,Kal2,Hol}).  For this to
apply, the dimension of $\LP$ as a lattice \ci{Kal1} (which is easily
seen to coincide with the dimension of $\P$ as a \tps) must be $\geq
4$, so that we must now assume that $\dim(\P)\neq 3$; the case
$\dim(\P)=2$ is covered directly by Axiom 2.  (The fact that dimension
3 is excluded is caused by the existence of so-called non-Desarguesian
projective geometries; see \ci{FH2} for a certain analogue of the
co-ordinatization procedure in that case.)

 Accordingly, for $\dim(\P)\neq 3$ there exists a vector space $V$
over a division ring $\Bbb D$ (both unique up to isomorphism),
equipped with an anisotropic Hermitian form $\th$ (defined relative to
an involution of $\Bbb D$, and unique up to scaling), such that
$\LP\simeq \L(V)$ as orthocomplemented lattices. Here $\L(V)$ is the
lattice of orthoclosed subspaces of $V$ (where the orthoclosure is
meant with respect to the orthogonality relation defined by $\th$).

 We shall now show that we can use Axiom 2 once again to prove that
${\Bbb D}={\Bbb C}$ as division rings. While this may seem obvious
from the fact \ci{FH1,Var} that for any irreducible projection lattice
one has ${\Bbb D}\simeq (\rh\vee \sg)\backslash \sg$ (for arbitrary
atoms $\rh\neq
\sg$), which is $\Bbb C$ by Axiom 2, this argument does not
prove that ${\Bbb D}=\C$ as division rings.

 The following insight (due to \ci{Kol}, and used in exactly the same
way in \ci{Zie} and \ci{CCR}) is clear from the explicit construction
of addition and mutliplication in $\Bbb D$ \ci{Var,FH1}.  Let $V$ be
3-dimensional, and let $\L(V)$ carry a topology for which the lattice
operations $\vee$ and $\wed$ are jointly continuous.  Then $\Bbb D$
(regarded as a subset of the collection of atoms in $\L(V)$), equipped
with the topology inherited from $\L(V)$, is a topological division
ring (i.e., addition and multiplication are jointly continuous).
 
Let $F\in\lp$ be finite-dimensional. We can define a topology on
$[\emptyset,F]$ (i.e., the set of all $Q\in\LP$ for which $Q\subseteq
F$) through a specification of convergence.

 Given a net $\{Q_{\lm}\}$ in $F$, we say that $Q_{\lm}\raw Q$ when
eventually $\dim(Q_{\lm})=\dim(Q)$, and if there exists a family of
bases $\{e_i^{\lm}\}$ for $\{Q_{\lm}\}$, and a basis $\{e_j\}$ of $Q$,
such that $\sum_{i,j}p(e_i^{\lm},e_j)\raw \dim(Q)$.  This notion is
actually independent of the choice of all bases involved, since
$\sum_jp(\rh,e_j)$ is independent of the choice of the basis in $Q$
for any $\rh\in\P$, and similarly for the bases of $Q_{\lm}$ (to see
this, extend $\{e_j\}_{j=1}^{\dim(Q)}$ to a basis
$\{e_j\}_{j=1}^{\dim(\P)}$, and use the property
$\sum_{j=1}^{\dim(\P)}p(e_j,\rh)=1$ for all $\rh\in\P$).

 An equivalent definition of this convergence is that $Q_{\lm}\raw Q$
if $p(\rh_{\lm},\sg)\raw 0$ for all $\sg\in F\wed Q^{\perp}$ and all
$\{\rh_{\lm}\}$ such that $\rh_{\lm}\in Q_{\lm}$.

Using the criteria in \ci{Kel}, it is easily verified that this
defines a topology on $F$. Moreover, this topology is Hausdorff. For
let $Q_{\lm}\raw Q$ and $Q_{\lm}\raw R$. Then $p(\rh_{\lm},\sg)\raw 0$
for all $\sg\in Q^{\perp}\vee R^{\perp}=(Q\wed R)^{\perp}$, and
$\{\rh_{\lm}\}$ as specified above.  Choose a basis $\{e_j\}$ of $Q$
which extends a basis of $Q\wed R$.  Then $\sum_{j=1}^{\dim(Q\wed
R)}p(\rh_{\lm},e_j)=1$, but also
$\sum_{j=1}^{\dim(Q)}p(\rh_{\lm},e_j)=1$ since $Q_{\lm}\raw Q$. Hence
$p(\rh_{\lm},\sg)\raw 0$ for all $\sg\in Q \wed (Q\wed R)^{\perp}$.
This leads to a contradiction unless $Q=R$.

{\bf Lemma 2} {\em The restriction of this topology to any two-sphere
$\rh\vee\sg\simeq S^2$ in $F$ induces the usual topology on $S^2$.
Moreover, $\vee$ and $\wed$ are jointly continuous on any
$[\emptyset,F]$, where $F$ is a 3-dimensional subspace of $\LP$. }

{\em Proof}. If we restrict this topology to the atoms in $F$, then
$\rh_{\lm}\raw
\rh$ if $p(\rh_{\lm},\rh)\raw 1$. This induces the usual topology on
$F=\sg\vee \ta \simeq S^2$, since one can easily show that, in
$F=\sg\vee \ta$, $p(\rh_{\lm},\rh)\raw 1$ is equivalent to
$p(\rh_{\lm},\nu)\raw p(\rh,\nu)$ for all $\nu\in\sg\vee\ta$ (cf.\
\ci{CCR}).

We now take $F$ to be a 3-dimensional subspace.  We firstly show that
$\rh_{\lm}\raw\rh$ and $\sg_{\lm}\raw\sg$, where $\rh$ and $\sg$ are
atoms, implies $\rh_{\lm}\vee\sg_{\lm}\raw\rh\vee\sg$. Let
$\ta_{\lm}=(\rh_{\lm}
\vee\sg_{\lm})^{\perp}\wed F$, and $\ta=(\rh\vee\sg)^{\perp}\wed F$;
 these are atoms.  Let $\rh_{\lm}'$ be the anti-podal point to
$\rh_{\lm}$ in $\rh_{\lm}\vee\sg_{\lm}$ (i.e.,
$\rh_{\lm}'=\rh_{\lm}^{\perp}\wed (\rh_{\lm}\vee\sg_{\lm})$), and let
$\sg_{\lm}'$ be the anti-podal to $\sg_{\lm}$ in $\rh_{\lm}
\vee\sg_{\lm}$.
Then $\{\rh_{\lm},\rh_{\lm}',\ta_{\lm}\}$ is a basis of $F$, and so is
$\{\sg_{\lm},\sg_{\lm}',\ta_{\lm}\}$.  The definition of a basis and
of $\rh_{\lm}\raw\rh$, $\sg_{\lm}\raw\sg$ implies that
$p(\rh,\ta_{\lm})\raw 0$ and $p(\sg,\ta_{\lm})\raw 0$. Hence
$p(\ta,\ta_{\lm})\raw 1$.  Now take an arbitrary atom
$\al_{\lm}\in\ta_{\lm}^{\perp}\wed F$, and complete to a basis
$\{\al_{\lm},\al_{\lm}',\ta_{\lm}\}$, where
$\al_{\lm}'\in\rh_{\lm}\vee\sg_{\lm}$.  Again, the definition of a
basis implies that $p(\al_{\lm},\ta)\raw 0$.  Hence by our second
definition of convergence $\rh_{\lm}\vee\sg_{\lm}\raw\rh\vee\sg$.

Secondly, we show that $Q_{\lm}\raw Q$ and $R_{\lm}\raw R$, where $Q$
and $R$ are two-dimensional subspaces of $F$, implies $Q_{\lm}\wed
R_{\lm}\raw Q\wed R$ (we assume $Q\neq R$, so eventually $Q_{\lm}\neq
R_{\lm}$).  Let $\al=Q^{\perp}\wed F$, $\bt=R^{\perp}\wed F$,
$\gm=Q\wed R$, and $\gm_{\lm}=Q_{\lm}\wed R_{\lm}$; as a simple
dimension count shows, these are all atoms.  By assumption,
$p(\gm_{\lm},\al)\raw 0$ and $p(\gm_{\lm},\bt)\raw 0$. Since
$(\al\cup\bt)^{\perp}=(\al\vee\bt)^{\perp}$ by definition of $\vee$,
and $(\al\vee\bt)$ is two-dimensional, $\gm$ is the only point in $F$
which is orthogonal to $\al$ and $\bt$.  Hence $p(\gm_{\lm},\gm)\raw
1$; if not, the assumption would be contradicted.  But this is
precisely the definition of $Q_{\lm}\wed R_{\lm}\raw Q\wed R$.\enp

From the classification of locally compact connected division rings
\ci{WZ} we conclude that ${\Bbb D}=\C$ as division rings; the ring
structure is entirely determined by the topology.  Moreover, Lemma 2
implies that the orthocomplementation is continuous on 3-dimensional
subspaces. If one inspects the way the involution of $\D$ is
constructed in the proof of the lattice co-ordinatization theorem, one
immediately infers that this involution (of $\C$ in our case) must
then be continuous as well. It can be shown that $\C$ only possesses
two continuous involutions: complex conjugation and the identity map
\ci{Var}. The latter cannot define a non-degenerate sesquilinear form
(so that, in particular, the lattice $\L(V)$ could not be
orthomodular). Hence one is left with complex conjugation, and $V$
must be a complex pre-Hilbert space.

The fact that $V$ is actually complete follows from the
orthomodularity of $\LP$ (hence of $\L(V)$).  The proof of this
statement is due to \ci{AA}; see (also cf.\ \ci[Thm.\ 11.9]{Kal2}, or
\ci[Thm.\ 21.4.1]{BC}). We will therefore write $V=\H$.

We conclude that $\lp$ is isomorphic to the lattice ${\cal L}(\H)$ of
closed subspaces of some complex Hilbert space $\H$.  Therefore, their
respective collections of atoms $\P$ and $\PH$ must be isomorphic.
Accordingly, we may identify $\P$ and $\PH$ as sets.  Denote the
standard \tpies\ (\ref{tpsforca}) on $\PH$ by $p_{\mbox{\tiny $\H$}}$.
With $p$ the \tpies\ in $\P$, we will show that $p=p_{\mbox{\tiny
$\H$}}$.
 
 Refer to the text following Definition \ref{deftwosphere}.  We may
embed $S_{\rm ref}^2$ isometrically in $\PH$; one then simply has
$p=p_{\mbox{\tiny $\H$}}$ on $S_{\rm ref}^2$.  Eq.\ (\ref{Tiso}) now
reads
\be
p_{\mbox{\tiny
$\H$}}(T_{\rh\vee\sg}(\rh'),T_{\rh\vee\sg}(\sg'))=p(\rh',\sg');
\ll{pmtiny}
\ee
in particular, $p_{\mbox{\tiny
$\H$}}(T_{\rh\vee\sg}(\rh'),T_{\rh\vee\sg}(\sg'))=0$ iff
$p(\rh',\sg')=0$.  On the other hand, we know that $p$ and
$p_{\mbox{\tiny $\H$}}$ generate isomorphic lattices, which implies
that $p_{\mbox{\tiny $\H$}}(\rh',\sg')=0$ iff $p(\rh',\sg')=0$.
Putting this together, we see that $p_{\mbox{\tiny
$\H$}}(T_{\rh\vee\sg}(\rh'),T_{\rh\vee\sg}(\sg'))=0$ iff
$p_{\mbox{\tiny $\H$}}(\rh',\sg')=0$.  A fairly deep generalization of
Wigner's theorem (see \ci[Thm.\ 4.29]{Var}; here the theorem is stated
for infinite-dimensional $\H$, but it is valid in finite dimensions as
well, for one can isometrically embed any finite-dimensional \Hs\ in
an infite-dimensional separable \Hs) states that a bijection
$T:\PH_1\raw\PH_2$ (where the $\H_i$ are separable) which merely
preserves orthogonality (i.e., $p_{\mbox{\tiny
$\H_2$}}(T(\rh'),T(\sg'))=0$ iff $p_{\mbox{\tiny
$\H_1$}}(\rh',\sg')=0$) is induced by a unitary or an anti-unitary
operator $U:\H_1\raw \H_2$. We use this with $\H_1=\rh\vee\sg$,
$\H_2=S_{\rm ref}^2$, and $T=T_{\rh\vee\sg}$. Since $T_{\rh\vee\sg}$
is induced by a(n) (anti-) unitary map, which preserves
$p_{\mbox{\tiny $\H$}}$, we conclude from (\ref{pmtiny}) that
$p_{\mbox{\tiny $\H$}}(\rh',\sg')=p(\rh',\sg')$.  Since $\rh$ and
$\sg$ (and $\rh',\sg'\in\rh\vee\sg$) were arbitrary, the proof of
Theorem \ref{chofp} is finished.\enp
\subsection{Transition probabilities}
Our aim is to show that the transition probabilities defined by
(\ref{mtp}) on the pure state space $\PA$ of the $C^*$-algebra
$\A_{\C}$ (i.e., $K=\SA$; recall that $\Ac$ is unital) coincide with
those originally defined on $\P=\PA=\eb$ (cf.\ Axiom 5); from
Proposition \ref{chofp} we know that these are given by
(\ref{tpsforca}).

 Firstly, $\A$ as a Banach space (and as an order-unit space) is
isomorphic to the space $A(K)$ of continuous affine functions on $K$,
equipped with the sup-norm.  The double dual $\A^{**}$ is isomorphic
to $A_b(K)$ (with sup-norm), and the $w^*$-topology on $A_b(K)$ as the
dual of $A(K)^*$ is the topology of pointwise convergence, cf.\
\ci{AE,AS1}. Since $A(K)$ is $w^*$-dense in $A_b(K)$, 
one may take the infimum in (\ref{mtp}) over all relevant $f$ in
$A(K)$.  Since $\A\subseteq \M\subseteq \A^{**}$ (where $\M$ was
defined in Proposition \ref{propapa}), by (\ref{apaispira}) one may
certainly take the infimum over $\AP$.  But, as we saw in section 3.4,
$\A_{00}(\P)$ is dense in $\AP$ when both are seen as subspaces of
$\ell^{\infty}(\P)$ with the topology of pointwise convergence.  Hence
we may take the infimum in (\ref{mtp}) over all relevant $f$ in
$\A_{00}(\P)$.

Let $Q$ be an orthoclosed subspace of $\P$, and recall that $p_Q$ was
defined in (\ref{defpqeq}).  We now show that an equation similar to
eq.\ (2.19) in \ci{AS1} holds, viz.\
\be
p_Q= {\rm inf}\, \{g\in \A_{00}(\P)|\, 0\leq g\leq 1, \,
g\upharpoonright Q=1\}. \ll{inf}
\ee
For suppose there exists a $0\leq h<p_Q$ for which the infimum is
reached.  We must have $h=1$ on $Q$ and $h=0$ on $Q^{\perp}$, since
$p_Q=0$ on $Q^{\perp}$.  Then the function $p_Q-h$ is $\geq 0$, and
vanishes on $Q$ and $Q^{\perp}$.  But such functions must vanish
identically: let $p_Q-h=\sum_i \lm_i\, p_{\rh_i}$.  Choose a basis
$\{e_j\}$ in $Q\cup Q^{\perp}$. For every point $\rh\in\P$, one must
have $\sum_j p(\rh,e_i)=1$. Hence $\sum_j (p_Q-h)(e_j)=\sum_i\lm_i=0$.
Suppose that $p_Q-h>0$.  Then there will exist another basis $\{u_j\}$
such that $f(u_j)>0$ for at least one $j$. This implies
$\sum_i\lm_i>0$, which contradicts the previous condition. We conclude
that $p_Q=h$, and (\ref{inf}) has been proved.

The desired result now follows immediately from (\ref{inf}) and the
observation that by definition $p_{\rh}(\sg)=p(\rh,\sg)$ for atoms
$Q=\rh$. \enp

We close this section with a technical comment. If $F\subset K \subset
\A^*$ (again with $K=\SA$) is a $w^*$-closed face, then $\partial_e F
\subseteq \partial_e K$ may be equipped with transition probabilities
defined by (\ref{mtp}), in which $A_b(K)$ is replaced by $A_b(F)$.
These coincide with the transition probabilities inherited from
$\partial_e K$. For $F=K \cap H$ for some $w^*$-closed hyperplane
$H\subset \A^*$ (see, e.g.,
\ci[II.5]{Alf1}, \ci[\S1]{AS1}), so that $A_b(F)\simeq H^*$.  By
Hahn-Banach, each element of $H^*$ can be extended to an element of
$\A^*$, so that any element of $A_b(F)$ extends to some element of
$A_b(K)$. The converse is obvious. The claim then follows from the
definition (\ref{mtp}).  This shows, in particular, that Axiom AHS2 is
equivalent to our Axiom 2.
\subsection{Poisson structure and orientability}
 While not necessary for the main argument in this paper, it is
enlightening to see that (given the other axioms) the existence of a
Poisson structure on $\P$ implies Axiom AHS5, i.e., orientability in
the sense of Alfsen et al.\
\ci{AHS} (also cf.\ \ci{Shu}). We still write $K$ for $\SA$.

 These authors define the object ${\cal B}(K)$ as the space of all
affine isomorphisms from $B^3$ onto a face of $K$ (which in our
setting is the state space of $\AP$ as a $JB$-algebra), equipped with
the topology of pointwise convergence.  It follows from Axiom 5 and
the argument in \ci[p.\ 499]{Shu} (or section 3 of \ci{CLM}) that one
can equally well work with the space ${\cal B}(\P)$ of all injective
maps from $S^2=\Bbb P\C^2$ into $\P$ which preserve transition
probabilities, topologized by pointwise convergence. If $\phv,\psi\in
{\cal B}(\P)$ have the same image, then by Axiom 2 and Wigner's
theorem the map $\psi^{-1}\circ\phv:S^2\raw S^2$ lies in $O(3)$
(acting on $S^2\in\R^3$ in the obvious way). The maps $\ps$ and $\phv$
are said to be equivalent if $\psi^{-1}\circ\phv\in SO(3)$; the space
of such equivalence classes is ${\cal B}(\P)/SO(3)$. The space $\P$ is
said to be orientable if the ${\Bbb Z}_2$-bundle ${\cal
B}(\P)/SO(3)\raw {\cal B}(\P)/O(3)$ is globally trivial (cf.\ \ci[\S
7]{AHS}). This notion of orientability is equivalent to the one used
in \ci{AHS}, cf.\
\ci{Shu}.

Given $\phv\in {\cal B}(\P)$ and $f\in\A$, we form $f\circ
\phv:S^2\raw \R$. We infer from the explicit description of
$\A$ in Chapter 3 that $f\circ \phv$ is smooth.  If $f,g\in\A$ then by
(\ref{hbarvaries})
\be
\{f,g\}\circ \phv(z)= \mbox{\rm sgn}(\phv) 
\hbar^{-1}(\phv(z))\{f\circ \phv,g\circ\phv\}_{S^2}(z),
\ll{pull}
 \ee where $\{\, ,\,\}_{S^2}$ is the Fubini-Study Poisson bracket on
$S^2$, and $\mbox{\rm sgn}(\phv)$ is $\pm 1$, depending on the
orientation of $\phv$.

 Now suppose that $K$ (hence $\P$) were not orientable. Then there
exists a continuous family $\{\phv_t\}_{t\in [0,1]}$ in ${\cal
B}(\P)$, for which $\phv_0$ and $\phv_1$ have the same image, but
opposite orientations (cf.\ the proof of Lemma 7.1 in \ci{AHS}, also
for the idea of the present proof). We replace $\phv$ by $\phv_t$ in
(\ref{pull}). Since $\{f,g\}$ is continuous, the left-hand side is
continuous in $t$ (pointwise in $z$). On the right-hand side,
$\{f\circ \phv_t,g\circ\phv_t\}_{S^2}$ is continuous in $t$, and so is
$\hbar^{-1}\circ\phv_t$. But $\mbox{\rm sgn}(\phv_t)$ must jump from
$\pm 1$ to $\mp 1$ between $0$ and 1, and we arrive at a
contradiction.

\end{document}